\documentclass[preprint,12pt,authoryear]{elsarticle}

\usepackage{amssymb}
\usepackage{amsmath}

\usepackage[utf8]{inputenc}
\usepackage[T1]{fontenc}
\usepackage{lineno}
\usepackage{url}
\usepackage[colorlinks=true,linkcolor=blue,citecolor=blue,urlcolor=blue]{hyperref}
\usepackage{placeins}
\usepackage{booktabs}

\journal{Icarus}
\begin{document}

\begin{frontmatter}

\title{Microphysical Model of Jupiter's Great Red Spot Upper Chromophore Haze} 

\author[label1,label2]{Asier Anguiano-Arteaga\corref{cor1}} 
\ead{asier.anguiano@ehu.eus}
\cortext[cor1]{Corresponding author.}

\author[label1]{Santiago Pérez-Hoyos} 
\author[label1]{Agustín Sánchez-Lavega} 
\author[label2]{Patrick G.J. Irwin} 

\affiliation[label1]{organization={Dpto. Física Aplicada, EIB, Universidad del País Vasco UPV/EHU},
            city={Bilbao},
            country={Spain}}
            
\affiliation[label2]{organization={Department of Physics, Atmospheric, Oceanic and Planetary Physics, University of Oxford},
	city={Oxford},
	country={United Kingdom}}

\begin{abstract}
The origin of the red colouration in Jupiter’s Great Red Spot (GRS) is a long--standing question in planetary science. While several candidate chromophores have been proposed, no clear conclusions have been reached regarding its nature, evolution, or relationship to atmospheric dynamics. In this work, we perform microphysical simulations of the reddish haze over the GRS and quantify the production rates and timescales required to sustain it. Matching the previously reported chromophore column mass and effective radius in the GRS requires column--integrated injection fluxes in the range $1\times10^{-12}$ -- $7\times10^{-12}$ kg m$^{-2}$ s$^{-1}$, under low upwelling velocities in the upper troposphere ($v_{\mathrm{trop}}\lesssim1.5\times10^{-4}$ m s$^{-1}$) and particle charges of at least 20 electrons/$\mu$m. Such rates exceed the mass flux that standard photochemical models of Jupiter currently supply via NH\textsubscript{3}--C\textsubscript{2}H\textsubscript{2} photochemistry at 0.1--0.2 bar, the most popular chromophore pathway in recent literature. We find a lower limit of 7 years on the haze formation time. We also assess commonly used size and vertical distribution parameterisations for the chromophore haze, finding that eddy diffusion prevents the long--term confinement of a thin layer and that the extinction is dominated by particles that can be represented by a single log--normal size distribution.
\end{abstract}



\end{frontmatter}


\section{Introduction}
\label{Introduction}
Jupiter’s Great Red Spot (GRS) is one of the most striking features in any planetary atmosphere. It is the largest anticyclone ever observed and also the longest--lived. Once thought to have been first observed by Giovanni Cassini in 1665, recent evidence indicates that the GRS was most likely first reported in 1831 and originated from a flow disturbance between the westward zonal jet at 20$^{\circ}$S and the eastward jet at 26$^{\circ}$S \citep{SanchezLavega2024}. Since the Voyager 1 and 2 Jupiter flybys \citep{Conrath1981,Flasar1981,Mitchell1981}, the GRS has been extensively studied. However, many aspects of its structure, internal dynamics, and composition remain uncertain.

Although Jupiter's colours are often exaggerated by image processing \citep{OrdonezEtxeberria2016,Irwin2024}, the Great Red Spot's remarkable reddish coloration is still one of its most debated aspects. The jovian condensible species predicted from thermochemical equilibrium models are white at visible wavelengths \citep{West1986,West2004}. These species cannot account for the broad blue--light absorption that is observed not only in the GRS, but over all the planet, including whitish regions such as the South Tropical Zone \citep{AnguianoArteaga2021}. The source of the jovian coloration is generally attributed to the optical properties and distribution of aerosols above the 1--bar pressure level \citep{Banfield1998}. A number of colouring aerosols, commonly referred to as chromophores, have been proposed, and a comprehensive review is given by \citet{West1986,West2004}. More recently, \citet{Loeffler2016} proposed irradiated ammonia hydrosulfide as a possible chromophore for the GRS, although its laboratory spectra show a distinct absorption band near 600 nm and comparatively weak absorption at wavelengths longer than 500 nm, features not seen in the GRS or other Jovian spectra \citep{Dahl2021}. 

The most popular chromophore candidate to date is the suite of products resulting from photolyzed ammonia (NH$_3$) reacting with acetylene (C$_2$H$_2$), as proposed by \citet{Carlson2016}, hereafter referred to as the ``Carlson chromophore''. These compounds, previously explored by \citet{FerrisIshikawa1987}, constitute an attractive candidate because acetylene is the third most abundant hydrocarbon species in Jupiter after methane and ethane, and ammonia is the fifth most abundant compound after hydrogen, helium, methane, and water \citep{SanchezLavega2011}. \citet{Carlson2016} did not identify a single specific compound responsible for the chromophore, but instead characterized a family of laboratory--produced residues with molecular formulae C$_{2n}$H$_{4n-2}$N$_2$, C$_{2n}$H$_{4n}$N$_2$ and C$_{2n}$H$_{4n+2}$N$_2$, with the H$_{4n}$ series being the most abundant. The associated wavelength-dependent imaginary refractive indices provided by \citet{Carlson2016}, corresponding to a 70 h irradiated sample, have been shown to reproduce a variety of observed Jovian spectra from different regions of the planet \citep{Sromovsky2017,Braude2020,PerezHoyos2020,Dahl2021,AnguianoArteaga2021,AnguianoArteaga2023,Fry2023}, including spectra specifically of the GRS \citep{Baines2019,Braude2020,AnguianoArteaga2021,AnguianoArteaga2023}, with little or no modification. A widely adopted chromophore configuration is the so--called Crème--Brûlée model, in which the chromophore is confined to a thin layer above the main tropospheric cloud. This configuration was first proposed for the Great Red Spot (GRS) by \citet{Baines2016}, and later extended to other regions across the planet. Notably, \citet{Sromovsky2017} suggested that the Carlson chromophore could act as a universal colouring agent throughout Jupiter’s atmosphere.

Although the spectral behaviour of the Carlson chromophore is plausible, the photochemical modelling of \citet{Moses2010} points to weak coupling between C$_2$H$_2$ and NH$_3$ in Jupiter’s troposphere as a result of the low diffusive flux at the tropopause. C$_2$H$_2$ is mainly produced by CH$_4$ photodissociation and reaches its maximum at very low pressures in the upper stratosphere (around $\sim 10^{-4}$~mbar), decreasing towards higher pressures \citep{Knizek2026}. In contrast, NH$_3$ is strongly depleted at pressures lower than $\sim 700$~mbar \citep{Moses2010}. Consequently, the vertical overlap between C$_2$H$_2$ and NH$_3$ is limited, and the weak diffusive transport across the tropopause inhibits efficient mixing of the two species. This issue was explored by \citet{Baines2019}, who estimated the diffusive flux of C$_2$H$_2$ into the high troposphere and concluded that an enhanced C$_2$H$_2$ supply would be required to generate the observed chromophore within plausible timescales (from $\sim$1.5 months to $\sim$11.5 years). As a possible source of this enhancement, \citet{Baines2019} invoked lightning--driven production of C$_2$H$_2$, as previously proposed for Jupiter by \citet{BarNun1985} and \citet{Podolak1988}. However, estimations made by \citet{Baines2019} assume a vertically constant eddy diffusion coefficient and neglect sedimentation, which becomes increasingly important with particle growth, since larger particles sediment more rapidly and would therefore remove material from the production region.

Building on this, a dedicated microphysical treatment is needed. Radiative transfer studies constrain the chromophore’s effective radius, column mass, and pressure level in the GRS, but they are not, on their own, sufficient to constrain the particle injection rates, vertical transport, and coagulation processes required to maintain such a layer. In particular, radiative transfer modelling is essentially steady--state: it constrains optical properties and assumed vertical distributions, but does not explicitly represent time--dependent evolution or rates. A one--dimensional microphysical model that explicitly includes sedimentation, eddy diffusion, and particle growth provides this link: while it remains an idealised representation of an inherently three--dimensional system, it allows us to estimate the evolution timescales and mass--injection rates needed to sustain the retrieved chromophore layer under physically plausible vertical velocities and particle sizes. Accordingly, rather than seeking a unique solution, we identify ranges of parameter combinations consistent with the available constraints and that reach steady state on timescales shorter than, or comparable to, a Jovian year.

In this work, we develop a one--dimensional microphysical model for the GRS chromophore haze and use it to quantify the required production rates and timescales. Section~\ref{micro_model} describes the model formulation, the adopted GRS atmospheric structure and eddy--diffusion profile, and the radiative transfer constraints on chromophore mass column density (MCD), particle size, and pressure level that we aim to reproduce for the \citet{Carlson2016} chromophore. We then define the parameter space explored. In Section~\ref{results}, we present the subset of simulations that match the retrieved chromophore properties in the GRS, derive the corresponding mass--injection rates and convergence timescales. In Section~\ref{discussion}, we interpret these results in terms of the availability of precursor gases (C$_2$H$_2$ and NH$_3$) and the resulting particle size and vertical distributions. Our main conclusions are summarised in Section~\ref{conclusions}. Finally, \ref{app1} validates the numerical implementation of the transport and coagulation schemes against analytical solutions.

\section{Microphysical model}
\label{micro_model}

\subsection{Haze microphysics}
\label{haze_micro}

We have developed a one--dimensional microphysical code based on the algorithms of \citet{Toon1988}. In Jupiter's upper atmosphere, the chromophore is expected to originate from gas--phase photochemistry, followed by conversion to particulate material. The code assumes that particles of a specified size are introduced through a prescribed net injection term, which parameterises the gas--to--particle conversion into particulate chromophore material, without explicitly modelling the underlying physico--chemical pathway (e.g., nucleation or polymerisation leading to solid particulate products). Once formed, particles are subject to sedimentation, eddy diffusion, and coagulation: sedimentation and eddy diffusion govern vertical transport, while coagulation alters the particle size distribution. Condensation and evaporation are not considered, following \citet{Carlson2016}, who indicate that their proposed chromophore behaves as a non--volatile solid residue. Using radiative transfer retrievals, \citet{Baines2019} investigated whether this chromophore could instead appear as a coating on the main tropospheric cloud particles, rather than as a distinct population of chromophore--only particles, but found that this assumption led to significantly poorer fits to Cassini VIMS spectra of the Great Red Spot (GRS). An alternative possibility is that the chromophore material acts as condensation nuclei for other tropospheric condensables; however, the resulting composite particles would modify the chromophore optical properties (in particular the imaginary refractive index) that have been shown to reproduce a wide range of Jovian spectra in radiative transfer studies, as previously discussed. Accordingly, we focus on the transport and coagulation of the chromophore--bearing haze once formed, following the parameterised source approaches adopted by \citet{Toon1980} in their study of Titan, and by \citet{Toledo2019} for Uranus and \citet{Toledo2020} for Neptune.

The aerosol continuity equation on which our model is based can be written as follows:

\begin{align}
	\frac{\partial C(z,r)}{\partial t} &= C_{\text{inj}}(z,r) \notag \\
	&\quad + \frac{\partial [ K_{zz}(z) \rho] }{\partial z}  \frac{\partial}{\partial z} \left[ \frac{C(z,r)}{\rho} \right]
	- \frac{\partial \left[ W(z,r)  C(z,r) \right]}{\partial z} \notag \\
	&\quad + P_{\text{coag}}(z,r) - L_{\text{coag}}(z,r)
	\label{eq:continuity}
\end{align}

Here, \( C(z,r) \) represents the number density of haze particles of radius \( r \) at altitude \( z \), and \( C_{\text{inj}}(z,r) \) denotes the injection rate of such particles at that level. The second and third terms on the right--hand side describe vertical transport: the former accounts for eddy diffusion, involving the eddy diffusion coefficient \( K_{zz}(z) \) and the atmospheric gas density \( \rho \), and the latter captures the effects of gravitational sedimentation and vertical advection via the vertical velocity \( W(z,r) \), which combines the prescribed background vertical motion with the particle settling velocity. Finally, the last two terms, $P_{\mathrm{coag}}(z,r)$ and $L_{\mathrm{coag}}(z,r)$, account for the net effect of coagulation on particles of radius $r$: production by aggregation of smaller particles, and loss due to coagulation with particles of the same or different sizes, respectively.

This formalism builds upon the one--dimensional model originally developed by \citet{Turco1979a} and \citet{Toon1979}. Various versions of this model have been widely used to study the atmospheres of Solar System bodies \citep{Pollack1987, Cabane1992, Moreno1996, Colaprete2003, McGouldrick2007, Toledo2020}, as well as those of giant exoplanets \citep{Marley2013, Gao2017}. Numerical validations of our code are presented in \ref{app1}.

\subsection{Atmospheric profiles in the GRS}
\label{atm_profiles}
\begin{figure}[ht]
	\centering
	\includegraphics[width=0.8\textwidth]{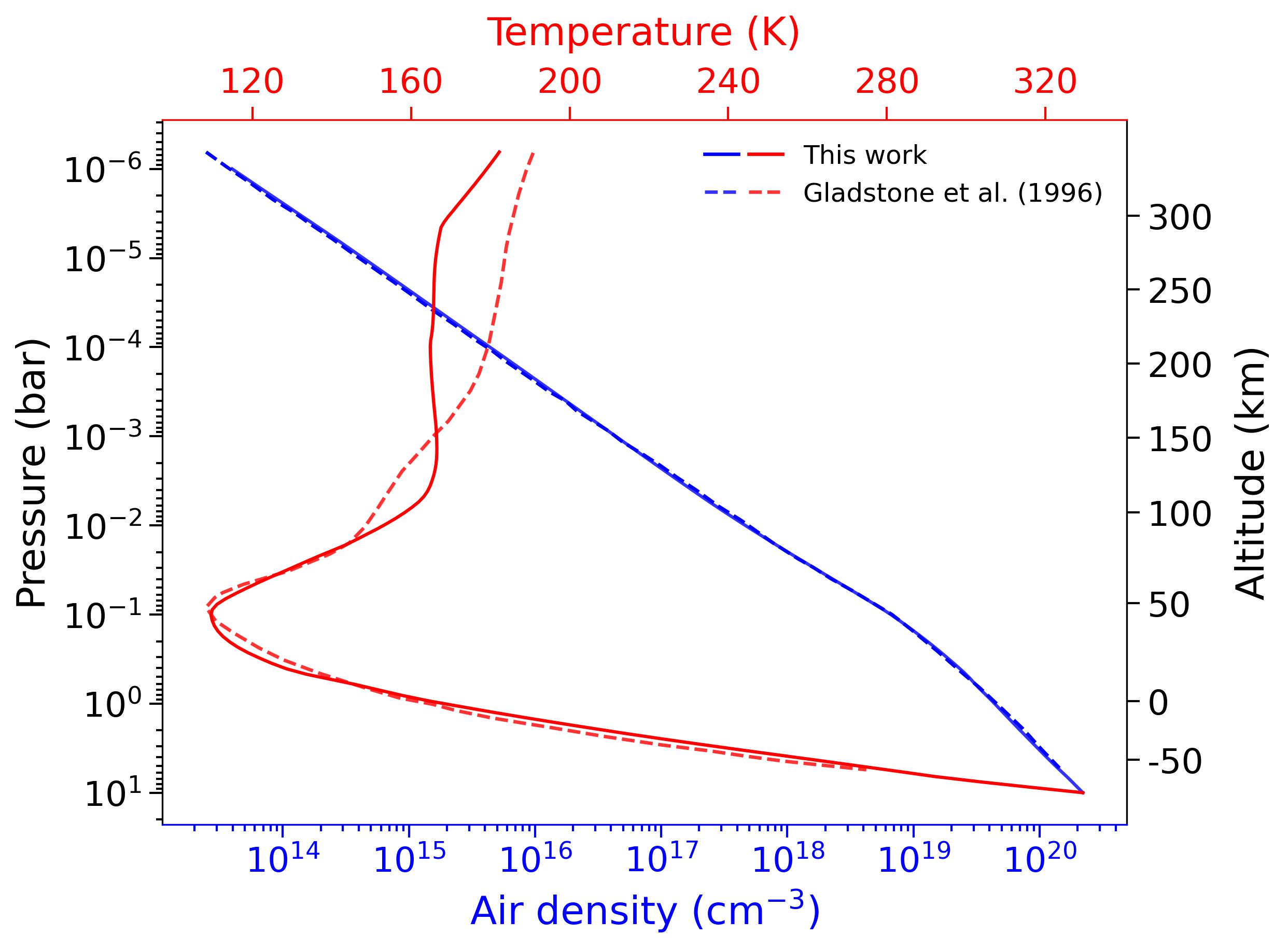}
	\caption{Profiles of temperature (red) and air number density (blue) as a function of pressure and altitude in the GRS (solid lines). The red and blue dashed lines show NEB profiles from \citet{Gladstone1996}, shown for comparison.}
	\label{fig:ptprofile}
\end{figure}

For the pressure--temperature (P--T) profile, we used the one derived from TEXES/Gemini North observations in March 2017, as published in the supplementary material of \citet{Fletcher2020}. Although that work focuses on the North Equatorial Belt (NEB), the same retrieval methodology was applied across a broad pressure range (10 to \(10^{-6}\) bar) and spatial domain (planetocentric latitudes \(0^\circ\) to \(26.5^\circ\)S, SIII west longitudes \(5^\circ\) to \(55^\circ\)), encompassing the location of the GRS at the time of the observations. To construct the P--T profile, we averaged temperatures over a \(3 \times 3\) pixel region within the GRS, based on a map with a spatial resolution of \(1^\circ\) per pixel in both latitude and longitude. From the resulting P--T profile, air density was computed using the ideal gas law, and pressure was converted to altitude via the hydrostatic equilibrium equation, assuming an effective gravitational acceleration of \(g = 23.17~\mathrm{m\,s^{-2}}\) at the GRS planetocentric latitude of \(20.5^\circ\)S. The resulting profile is shown in Fig.~\ref{fig:ptprofile}. As a basic consistency check, we compared these profiles with those presented by \citet{Gladstone1996}, which provide self--consistent thermal and air density vertical profiles, to verify that our temperature and air density values are sound. We found good overall agreement. The remaining differences, most noticeably in temperature at pressures lower than $10^{-2}$~bar, are reasonable given that the \citet{Gladstone1996} profiles correspond to a different region (the NEB) and a different epoch.

\begin{figure}[ht]
	\centering
	\includegraphics[width=0.8\textwidth]{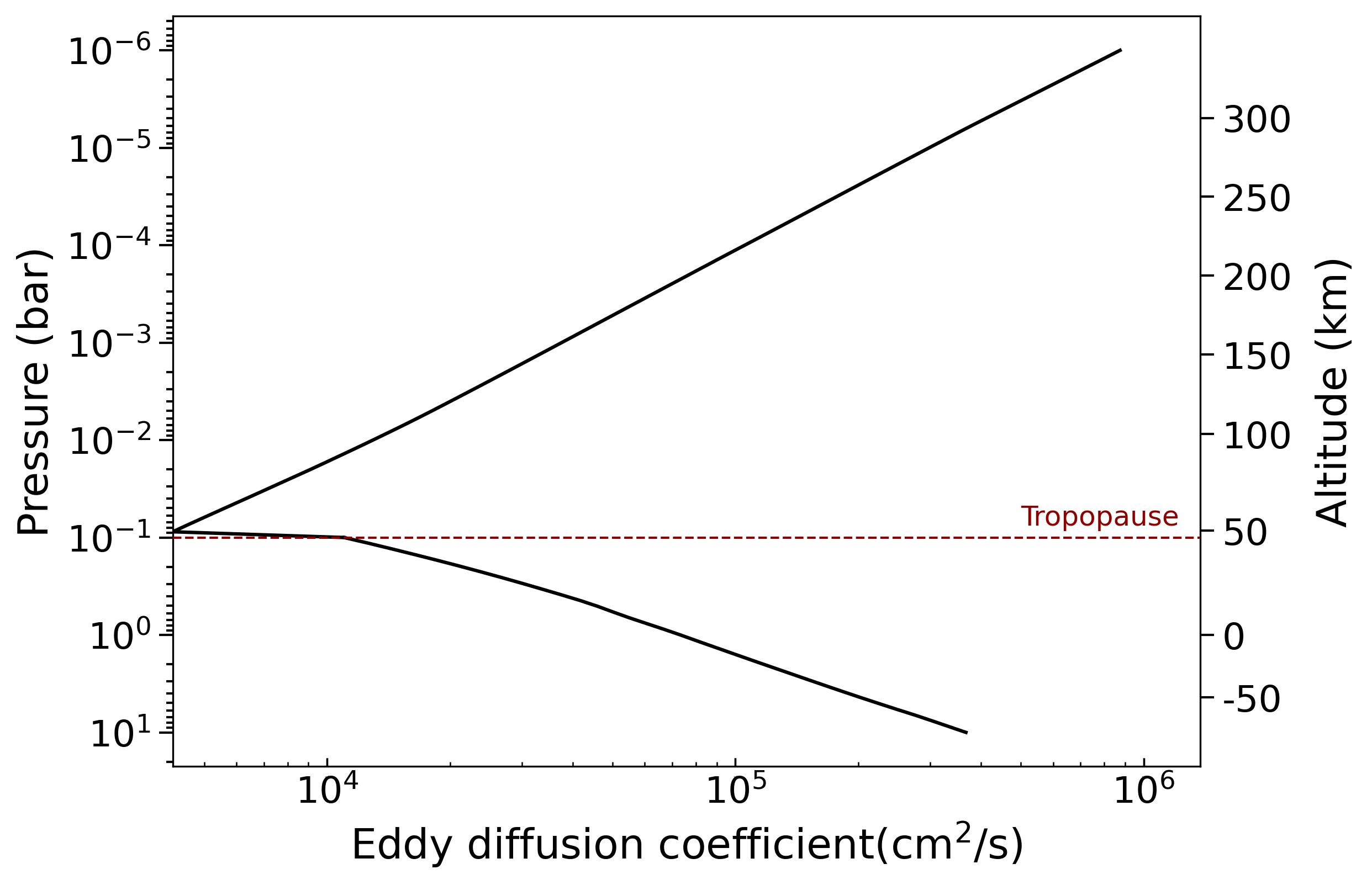} 
	\caption{
		Vertical profile of the eddy diffusion coefficient used for the GRS.}
	\label{fig:Kprofile}
\end{figure}

For the vertical profile of eddy diffusion $K_{zz}(z)$ shown in Eq.~\ref{eq:continuity}, we adopted the formulation proposed by \citet{Irwin2009}:

\begin{equation}
	K_{zz}(z) = 
	\begin{cases}
		K_{zz}(z_H) \left[ \dfrac{N_H}{N(z)} \right]^\gamma, & \text{for } z \geq z_T \\[12pt]
		K_{zz}(z_T) \left[ \dfrac{N(z)}{N_T} \right], & \text{for } z < z_T
	\end{cases}
	\label{eq:Kprofile}
\end{equation}
where \( z_T \) is the altitude of the tropopause (taken here to be at 0.1 bar), \( N(z) \) is the air number density, \( N_H \) is the air number density at the homopause (defined as the altitude below which eddy mixing dominates over molecular diffusion), with associated eddy diffusion coefficient \(K_{zz}(z_H)\). \( N_T \) is the air number density at the tropopause, \(K_{zz}(z_T)\) is the eddy diffusion coefficient at the tropopause, and $\gamma$ is a dimensionless exponent (typically $\sim 0.5$) that controls how rapidly $K_{zz}$ varies with air number density $N(z)$, consistent with gravity--wave--driven mixing scalings discussed by \citet{Irwin2009}. This formulation provides a $K_{zz}(z)$ profile that can be applied using the GRS--specific air number density profile, in the absence of a well--constrained eddy diffusion profile for this region, and is valid over our model domain ($z\simeq 0$--158~km; $P\simeq 1$--$10^{-3}$~bar). The vertical eddy diffusion profile used in this work is shown in Fig.~\ref{fig:Kprofile}, computed assuming \( K_{zz}(z_H) = 1.4 \times 10^6~\mathrm{cm^2\,s^{-1}} \) \citep{Irwin2009, Gladstone1996}, \( K_{zz}(z_T) = 1.1 \times 10^4~\mathrm{cm^2\,s^{-1}} \) \citep{Moses2005}, \( N_H = 1.4 \times 10^{13}~\mathrm{cm^{-3}} \) \citep{Irwin2009, Gladstone1996}, \( N_T = 6.59 \times 10^{18}~\mathrm{cm^{-3}} \) (our calculated value at the tropopause, i.e., at 0.1 bar), and \( \gamma = 0.45 \) \citep{Gladstone1996}.

According to \citet{Kasten1968}, the sedimentation velocity \( v_{\text{fall}} \) of a spherical particle with radius \( r \) and density \( \rho_p \), falling through a medium with gravitational acceleration \( g \) and dynamic viscosity \( \eta \), is given by:

\begin{equation}
	v_{\text{fall}} = \frac{2}{9} \cdot \frac{\rho_p r^2 g}{\eta} 
	\left[1 + \text{A}K_n + \text{B}K_n e^{-\text{C}/K_n}\right]
	\label{eq:vsed}
\end{equation}

\begin{figure}[ht]
	\centering
	\includegraphics[width=1.0\textwidth]{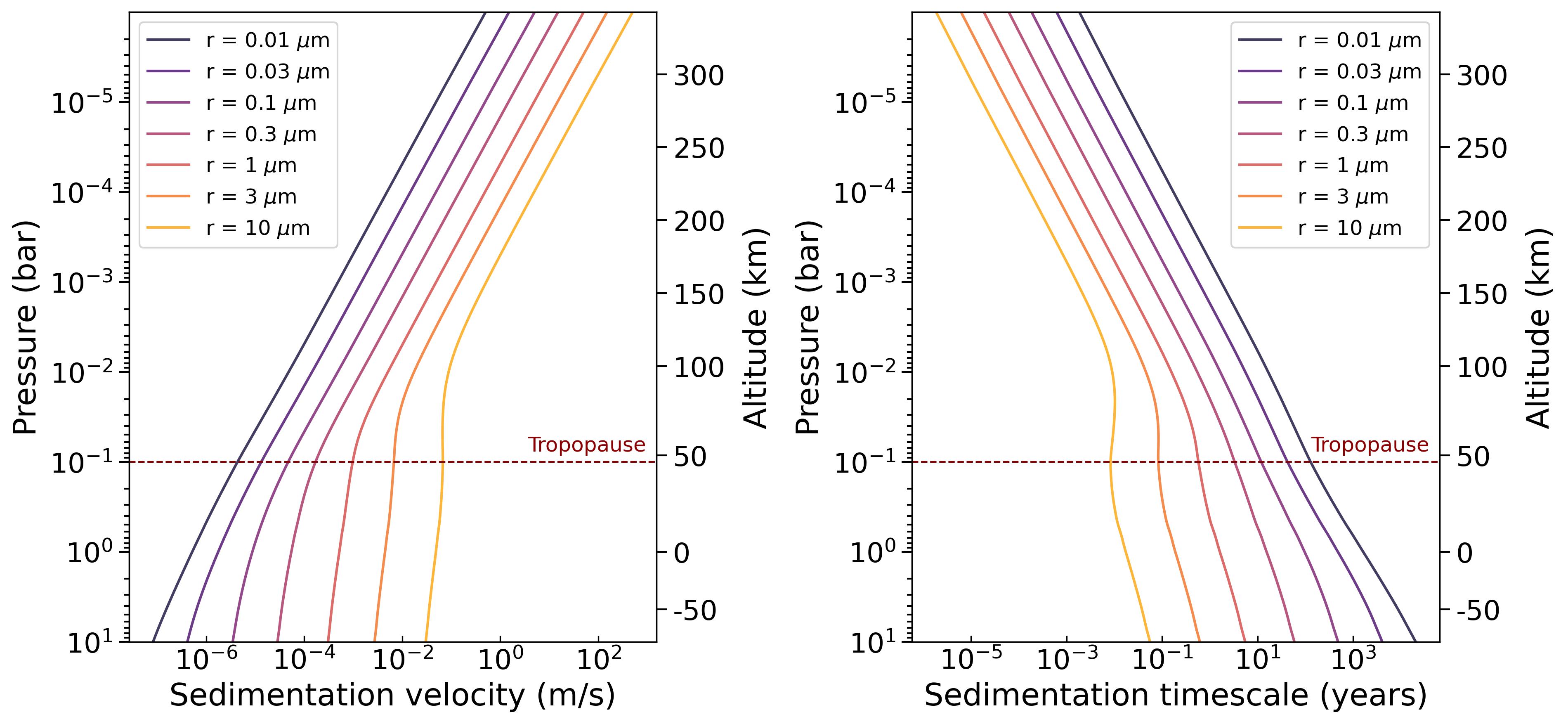} 
	\caption{Sedimentation velocities and timescales for particles with radii in the range 0.01--10~\(\mu\)m in the GRS according to the formulation of \citet{Kasten1968}.}
	\label{fig:vsed_profile}
\end{figure}

Here, \( K_n \) is the Knudsen number, a dimensionless parameter that characterizes whether the gas--particle interaction is in the continuum regime (\( K_n \ll 1 \)) or in the molecular (free molecular) regime (\( K_n \gg 1 \)) described by Maxwell--Boltzmann statistics. It is defined as
\begin{equation}
	K_n = \frac{\lambda_g}{r},
\end{equation}
where \( \lambda_g \) is the gas mean free path and \( r \) is the particle radius. The sedimentation velocity formulation in Eq.~\ref{eq:vsed} is valid across the full range of Knudsen numbers, smoothly bridging the continuum and free molecular regimes. To characterize the dynamic viscosity $\eta$ in the Jovian atmosphere, we adopted the profile given by \citet{Hansen1979}, which assumes a composition of 89$\%$ H$_{2}$ and 11$\%$ He and depends only on temperature. The correction factor in brackets accounts for slip--flow effects through a semi--empirical expression involving dimensionless coefficients \(A\), \(B\), and \(C\). We adopt the values \(A=1.257\), \(B=0.4\), and \(C=1.1\), based on the general formulation of \citet{Davies1945} for spherical particles moving through a background gas. \citet{Pruppacher2010} notes that these coefficients are widely recommended. Thus, the dependence of the sedimentation velocity on the Jovian background gas enters through the Knudsen number \(K_n\) and the dynamic viscosity \(\eta\). Figure~\ref{fig:vsed_profile} shows the sedimentation velocities and timescales for particles with radii between 0.01 and 10~\(\mu\)m in the GRS, assuming a particle density of \( \rho_p = 0.7~\mathrm{g/cm^3} \) (see Section~\ref{model_params} for our choice of this value). For particles with radius \( r = 0.1~\mu\mathrm{m} \), the sedimentation timescale at 0.1 bar, defined as \(\tau_{\rm fall} = H / v_{\rm fall}\) with $H$ the local scale height, is approximately \(12~\mathrm{years}\), comparable to a Jovian year.

\begin{figure}[ht]
	\centering
	\includegraphics[width=0.8\textwidth]{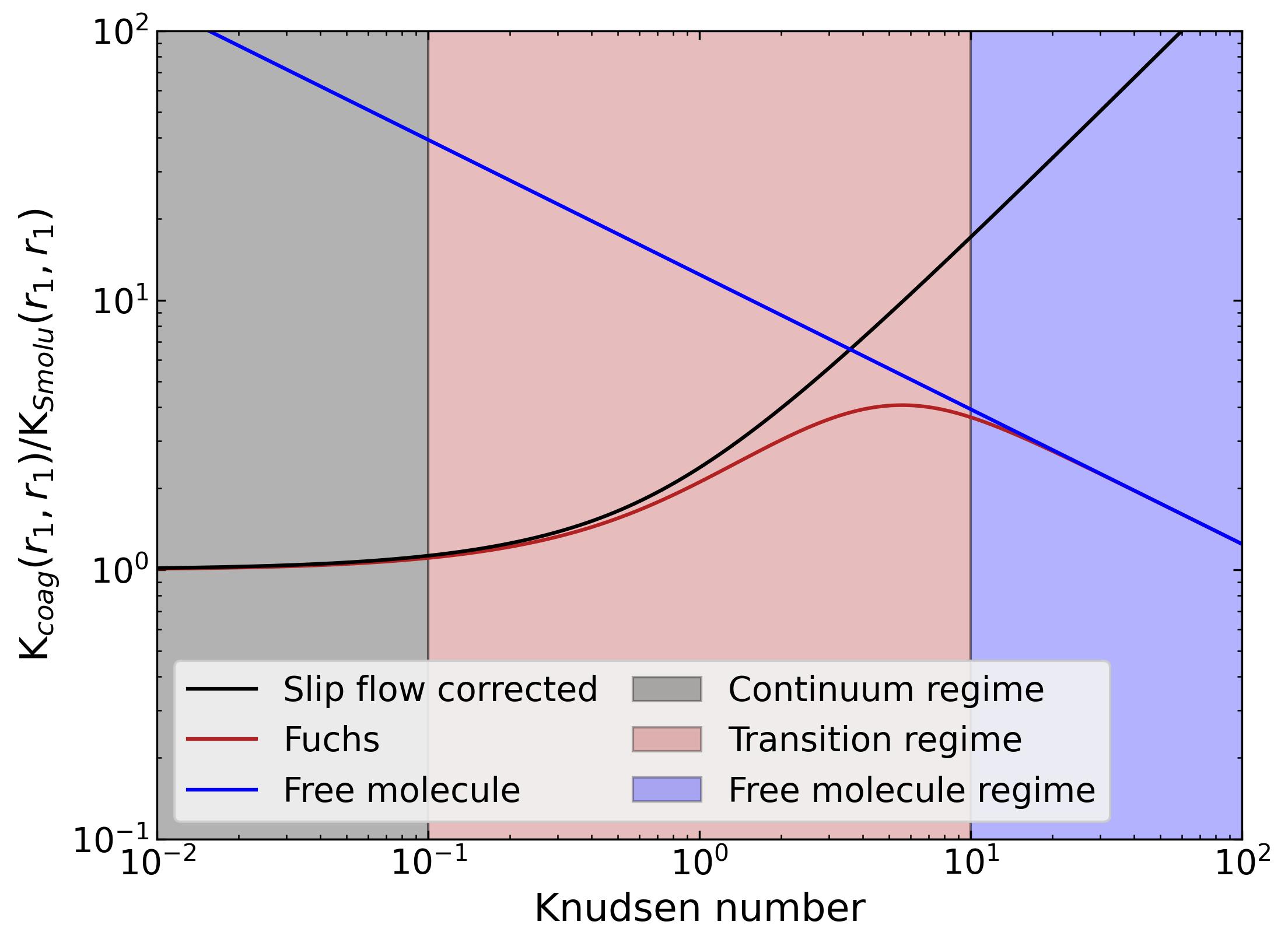}
	\caption{
		Ratio between the coagulation kernel \( K_{\text{coag}}(r_1, r_1) \) and the classical Smoluchowski constant \( K_{\text{Smolu}}(r_1, r_1) \) as a function of the Knudsen number, computed under terrestrial conditions (\( T = 298~\mathrm{K} \), \( \eta = 1.85 \times 10^{-5}~\mathrm{Pa\cdot s} \), mean free path \( \lambda_g = 6.86 \times 10^{-8}~\mathrm{m} \)).The black line corresponds to the continuum regime with slip--flow corrections, the red line represents the Fuchs formulation, and the blue line indicates the free--molecule limit. Shaded regions highlight the continuum, transition, and free molecular regimes.
	}
	\label{fig:coag_kernel_regimes}
\end{figure}

The coagulation production and loss terms, \( P_{\text{coag}} \) and \( L_{\text{coag}} \), in Eq.~\ref{eq:continuity} are governed by the coagulation kernel \( K_{\rm coag}(r_1, r_2) \) \citep{Toon1988}, which quantifies the rate at which particles of sizes \( r_1 \) and \( r_2 \) collide and stick together as a consequence of random thermal Brownian motion. In this work, we adopt the formulation of \citet{Sitarski1977}, originally developed by \citet{Fuchs1964}. Given the length of the full expression for the coagulation kernel, we refer the reader to \citet{Sitarski1977} for the detailed formulation. Its key advantage is its applicability across all flow regimes—continuum, transition, and free molecular. This is illustrated in Figure~\ref{fig:coag_kernel_regimes}, in which the ratio of the Fuchs coagulation kernel to the Smoluchowski coagulation constant \citep{Sitarski1977} is shown as a function of the Knudsen number.

\begin{figure}[ht]
	\centering
	\includegraphics[width=0.75\textwidth]{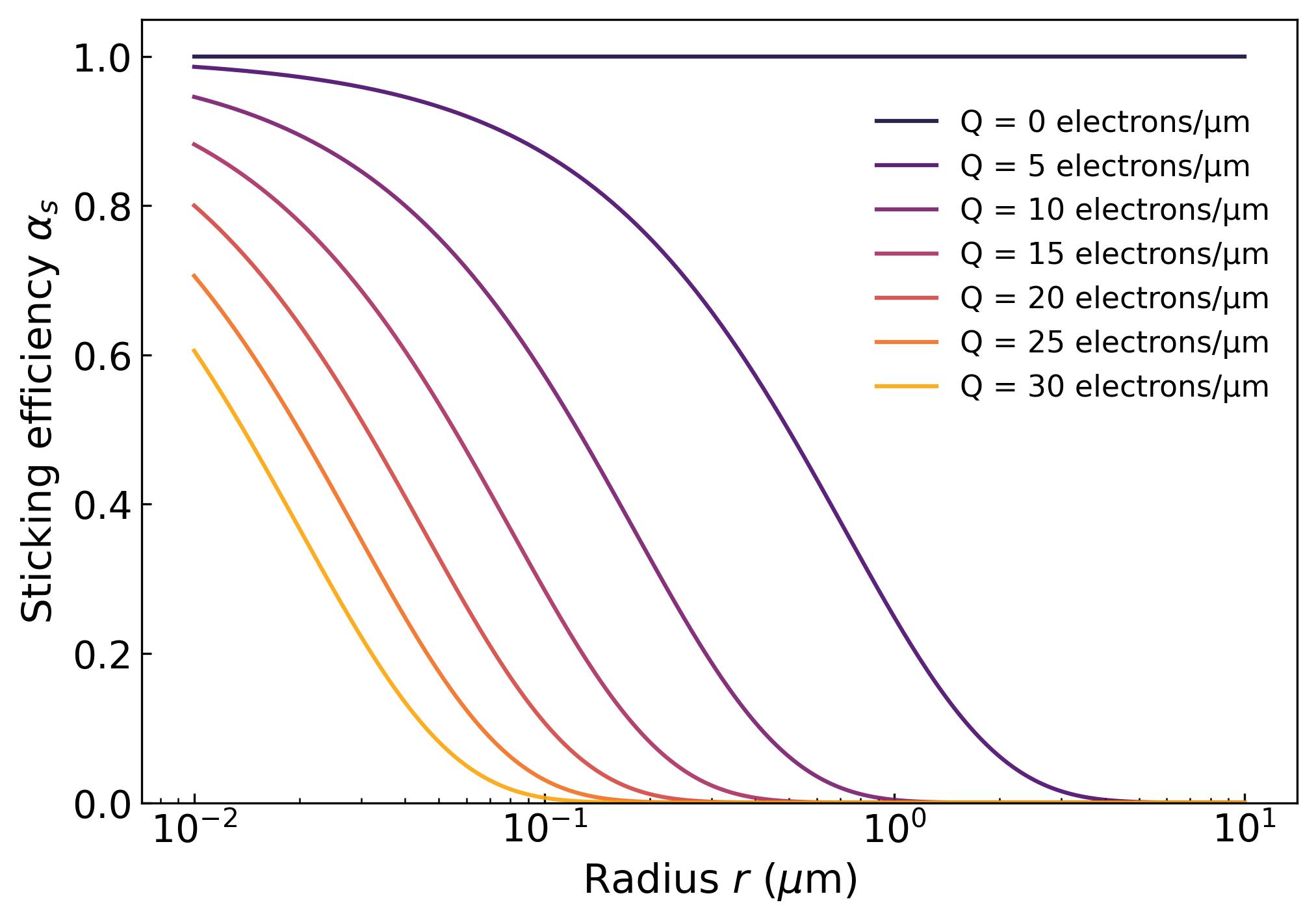}
	\caption{Sticking efficiency $\alpha_s$ (the multiplicative factor applied to the coagulation kernel) at $T=150$~K for collisions between equal--sized particles ($r_1=r_2=r$) as a function of particle radius, shown for the range of $Q$ values explored in this work. Note that, although equal--size coagulation can be strongly inhibited as Q increases, growth can still occur via collisions with smaller particles for which $\alpha_s(r_1,r_2)$ is not strongly suppressed.}
	\label{fig:sticking_efficiency}
\end{figure}

For charged particles, the coagulation kernel is scaled by the sticking efficiency \(\alpha_s\), defined as \citep{Pollack1987}:
\begin{equation}
	\alpha_s = \exp\left( \frac{-k_C Q^2 e^2 r_1 r_2}{k_B T (r_1 + r_2)} \right),
\end{equation}
where \(k_C = 1/(4\pi\varepsilon_0)\) is the Coulomb constant, \(Q\) is the number of elementary charges \(e\) per micron of radius, \(r_1\) and \(r_2\) the radii of the interacting particles, \(k_B\) the Boltzmann constant, and \(T\) the atmospheric temperature.  This coefficient accounts for the electrostatic repulsion between like--charged particles, which reduces their probability of coagulating.  Although explicit mention of the Coulomb constant \(k_C\) is typically absent from earlier formulations, its inclusion is required by dimensional analysis. This has been validated by comparison with the sticking efficiencies reported by \citet{Toon1980} for the cases of single--charged particles of size \(10^{-3}\), \(10^{-2}\), and \(10^{-1}\) \(\mu\)m. Figure~\ref{fig:sticking_efficiency} shows the sticking efficiency $\alpha_s$ for equal--sized particles ($r_1=r_2=r$) as a function of particle radius for the range of $Q$ values explored in this work (see Section~\ref{model_params}).

\subsection{Constraints from Radiative Transfer modelling}
\label{RT_constraints}
In our microphysical simulations, we aim to reproduce the values derived for the GRS, particularly the MCD and the effective radius of the chromophore particles. These quantities have been constrained in previous studies, including \citet{Baines2019}, \citet{Braude2020}, and \citet{AnguianoArteaga2021, AnguianoArteaga2023}. Significantly, there is an excellent agreement between the estimates of \citet{Baines2019} and \citet{AnguianoArteaga2021} for the chromophore MCD in the GRS, both reporting values around \(30~\mu\mathrm{g}/\mathrm{cm}^2\). A very similar value is found in the second preferred model of \citet{Braude2020}, which yields effective radii more consistent with those from the other two studies. In that model, the chromophore effective radius is fixed at $0.2~\mu$m based on their NEB limb--darkening analysis, resulting in a chromophore MCD of approximately \(30~\mu\mathrm{g}/\mathrm{cm}^2\). Across all these works focused on the GRS, the retrieved effective radii for the chromophore are of the order of a few tenths of a micron. This size range is likewise typical in the other radiative transfer studies discussed above.

We will therefore focus on the simulations showing a MCD in the uncertainty range constrained by the minimum value of \(20~\mu\mathrm{g/cm}^2\) reported by \citet{AnguianoArteaga2021} for the GRS nucleus and the maximum of \(40~\mu\mathrm{g/cm}^2\) obtained by \citet{Baines2019} for a vertically extended haze layer. As for the effective radius, we will consider simulations resulting in the range \(0.2\text{--}0.35~\mu\mathrm{m}\) from the values reported by \citet{AnguianoArteaga2023}. Notably, this study includes data from UV filters, which are particularly sensitive to particles of that size and to haze altitudes at the expected chromophore pressure levels (\citealt{PerezHoyos2012}, \citealt{SanchezLavega2013}). Both the MCD and the effective radius will be computed for pressures lower than 0.2~bar, corresponding to the chromophore pressure level reported by \citet{Sromovsky2017}, \citet{Baines2019}, and \citet{Braude2020}. This range also includes the base pressure level of the chromophore layer, around 0.1~bar, as indicated by \citet{AnguianoArteaga2021}.

\subsection{Model parameters}
\label{model_params}
To reproduce the retrieved properties of the chromophore layer in the GRS, we adopt a simplified yet physically motivated parameterisation of the microphysical model. The model is integrated on a one--dimensional vertical grid spanning pressures from 1 to $10^{-3}$~bar (0--160~km), discretised into 52 layers approximately equally spaced in log--pressure. Particles are injected at a constant rate following a Gaussian distribution in pressure \citep{Toon1980,Moreno1996}, for which we explore different Full Width at Half Maximum (FWHM) values.
In pressure coordinates, the vertical profile of the injection rate is
prescribed as
\begin{equation}
	C_{\mathrm{inj}}(P) = C_{\mathrm{inj}}
	\exp\!\left[-\frac{(P - P_{\mathrm{inj}})^2}{2\sigma^2}\right],
	\label{eq:Cinj_gauss}
\end{equation}
where \(P\) is the linear pressure, \(P_{\mathrm{inj}}\) is the peak injection pressure, and \(C_{\mathrm{inj}}\) denotes the peak injection rate (at \(P_{\mathrm{inj}}\)). The standard deviation \(\sigma\) is related to the FWHM via
\begin{equation}
	\sigma = \frac{\mathrm{FWHM}}{2\sqrt{2\ln 2}}.
	\label{eq:Cinj_FWHM}
\end{equation}
This injection, represented by the source term \(C_{\text{inj}}(z,r)\) in Eq.~\eqref{eq:continuity}, is treated as local production of chromophore particles, with no explicit inflow from the domain boundaries. The injected particle size is fixed at \(0.01~\mu\mathrm{m}\), following \citet{Pollack1987} and as done by \citet{Moreno1996} for the Jovian atmosphere. Comparable 1--D aerosol microphysical frameworks also adopt initial radii between \(0.001--0.002\)\(~\mu\mathrm{m}\) \citep[e.g.,][]{Toon1980,Toledo2019}. \citet{Pollack1987} noted that their smallest size bin corresponds to clusters comprising tens of molecules when molecular sizes are of the order of angstroms. This motivates our choice to keep the minimum radius safely above the molecular scale. Individual molecules of the chromophore family identified by \citet{Carlson2016} have characteristic sizes of several angstroms, and in some cases (e.g., aromatic rings or extended hydrocarbon chains) can approach nanometre scales. Adopting a minimum particle radius of \(0.01~\mu\mathrm{m}\) therefore ensures that the smallest size bin represents molecular clusters large enough that the particle geometry can be reasonably approximated as spherical.

A constant upward vertical velocity \(v_\mathrm{trop}\), consistent with lower--limit estimates reported by \citet{Conrath1981}, is imposed throughout the troposphere (i.e., the same value at pressures equal to or greater than 0.1~bar) and is added to the particle sedimentation velocity to form the vertical velocity \(W(z,r)=v_{\rm fall}(z,r)+v_\mathrm{trop}\) in Eq.~\ref{eq:continuity}. This term is included to represent the mean tropospheric upwelling in an effective 1--D column and is not intended to capture the full secondary circulation of the GRS, which comprises both upwelling and subsidence and is not uniquely constrained by an accepted dynamical model. Thus, our approach provides a reasonable approximation following the estimates of \citet{Conrath1981}, who inferred vertical velocities near the tropopause by assuming that local temperature departures are maintained by a balance between adiabatic heating/cooling associated with vertical motion and thermal damping. This prescription does not imply the absence of deposition: $v_\mathrm{trop}$ adds an advective contribution to the vertical transport and can reduce, balance, or even reverse the net downward motion from gravitational settling depending on the relative magnitudes of $v_\mathrm{trop}$ and $v_{\rm fall}$.

The particle material density is assumed to be 0.7~g/cm$^3$, using the bulk density of HCN reported by \citet{Haynes2011} as a representative value for organic nitrogen--bearing compounds, and consistent with the hydrocarbon particle densities considered by \citet{Pollack1987} and \citet{Toledo2019}.

To ensure numerical stability, we use an implicit time integration scheme, which is not limited by the Courant--Friedrichs--Lewy stability condition that constrains explicit schemes \citep{Press1992}. We adopt a fixed integration time step of 10$^4$~s, consistent with the initial value used by \citet{Pollack1987}. Numerical integrity is verified during the simulations by (i) monitoring for the appearance of negative particle number concentrations (in any size bin and altitude layer), which are non--physical, (ii) ensuring the absence of spurious numerical oscillations, and (iii) tracking the system's total mass to verify mass conservation. In particular, at any time the total mass within the domain equals the cumulative injected mass minus the mass lost through the lower boundary (downward sedimentation out of the model). All this allows for computationally efficient simulations over physically meaningful integration times. The simulations are run over a total integration time of almost 15 simulated Earth years, slightly more than one Jovian year, after which the effective radius and the MCD are evaluated for convergence towards a steady state. Convergence is assessed by requiring that both the effective radius and the MCD vary by less than 0.5\% over five consecutive outputs sampled every 1,000 time steps.

In order to constrain the microphysical parameters of the chromophore layer in the GRS, we explored a comprehensive grid of 2,160 models defined by five free parameters: the peak particle injection rate \(C_\mathrm{inj}\), the upward tropospheric velocity \(v_\mathrm{trop}\), the injection pressure level \(P_\mathrm{inj}\), the FWHM of the injection profile, and the number of electrons per micron of radius \(Q\). The explored ranges and sampling for each parameter are summarised in Table~\ref{tab:param_grid}.

The vertical velocity range, \(v_\mathrm{trop} \in [5.0 \times 10^{-5},\ 1.0 \times 10^{-2}]\) m s$^{-1}$ , was chosen based on dynamical constraints. The lower limit is motivated by the estimate of \(v \geq 4.5 \times 10^{-5}\) m s$^{-1}$ at the GRS tropopause by \citet{Conrath1981},  while the upper limit corresponds to a conservative value: the velocity required to fully suspend particles with radii up to \(3~\mu\mathrm{m}\) (such large particles are not expected to be efficiently suspended at upper tropospheric levels), which results in long convergence times exceeding 20 years. Within this range, six values were explored, evenly spaced in logarithmic scale.

Particle injection profiles (Eq.~\ref{eq:Cinj_gauss}) were centred at two pressure levels, 0.1 and 0.2~bar, following the chromophore layer locations inferred by \citet{Baines2019}, \citet{Braude2020}, and \citet{AnguianoArteaga2021, AnguianoArteaga2023}. Higher injection levels were excluded, as NH$_3$ (a key precursor in the Carlson chromophore pathway) is significantly depleted above these pressures. The width of the injection profile, defined by its FWHM (Eq.~\ref{eq:Cinj_FWHM}), was varied between 30\%, 60\%, and 90\% of the peak injection pressure. This choice spans relatively narrow and broad cases while avoiding excessive stratospheric injection. Expressing the FWHM as a fraction of the peak injection pressure allows the vertically integrated injection rate to remain comparable between profiles centred at 0.1 and 0.2~bar for a given peak value of \(C_\mathrm{inj}\). As shown in Section~\ref{results}, acceptable solutions are obtained for both injection pressures (\(P_{\rm inj}=0.1\) and \(0.2\)~bar) and for all FWHM values considered. As expected, broader injection profiles yield a larger vertically integrated (column) injection for a fixed peak injection rate \(C_{\rm inj}\); consequently, acceptable cases with smaller peak injection rates tend to require broader profiles to meet the target MCD, whereas cases with larger peak injection rates generally favour narrower profiles. To facilitate comparison across different injection widths and peak rates, we also discuss the source strength in terms of the associated column mass injection flux (in units of \(\mathrm{kg\,m^{-2}\,s^{-1}}\)), which provides a direct measure of the vertically integrated source.

The peak particle injection rate spanned from \(7.0\, \times \, 10^{-3}\) to \(10~\mathrm{particles\,cm^{-3}\,s^{-1}}\). The lower bound corresponds to a scenario where a FWHM of 90\% of \(P_\mathrm{inj}\) yields a total MCD of 20~\(\mu\mathrm{g/cm}^2\) over the simulation time, matching our lower acceptable limit. The upper bound leads to 30~\(\mu\mathrm{g/cm}^2\) being injected in just 20 days, ensuring that fast--growing scenarios are also captured. A total of ten values were explored, distributed logarithmically within the range.

Finally, particle charge was varied between 5 and 30 electrons per micron of radius, in increments of 5. This range was selected based on an extensive preliminary analysis of more than a thousand simulations indicating that values above \(Q = 0~ \text{electrons}/\mu\mathrm{m}\) were required to avoid excessively large effective radii at the target mass column densities. The upper limit of \(Q = 30~ \text{electrons}/\mu\mathrm{m}\) is based on the value reported by \citet{Moreno1996} for polar latitudes.

\begin{table}[t]
	\centering
	\caption{Summary of the explored parameter space (total: $2,160$ models). For $C_{\mathrm{inj}}$ and $v_{\mathrm{trop}}$, brackets indicate the lower and upper bounds of the explored ranges.}
	\label{tab:param_grid}
	\resizebox{\linewidth}{!}{%
		\begin{tabular}{@{}llll@{}}
			\toprule
			Parameter & Symbol & Explored values & Sampling \\
			\midrule
			Peak injection rate &
			$C_{\mathrm{inj}}$ &
			$[7.0\times 10^{-3},\ 10]~\mathrm{particles~cm^{-3}~s^{-1}}$ &
			$N=10$, log spaced \\
			
			Upward tropospheric velocity &
			$v_{\mathrm{trop}}$ &
			$[5.0\times 10^{-5},\ 1.0\times 10^{-2}]~\mathrm{m~s^{-1}}$ &
			$N=6$, log spaced \\
			
			Injection pressure level &
			$P_{\mathrm{inj}}$ &
			$\{0.1,\ 0.2\}~\mathrm{bar}$ &
			$N=2$ \\
			
			Injection profile width &
			$\mathrm{FWHM}$ &
			$\{0.3,\ 0.6,\ 0.9\}\,P_{\mathrm{inj}}$ &
			$N=3$, as a fraction of $P_{\mathrm{inj}}$ \\
			
			Particle charge per radius &
			$Q$ &
			$\{5,10,15,20,25,30\}~\mathrm{electrons\,\mu m^{-1}}$ &
			$N=6$ \\
			\bottomrule
		\end{tabular}%
	}
\end{table}

\section{Results}
\label{results}
\begin{figure}[h]
	\centering
	\includegraphics[width=0.7\textwidth]{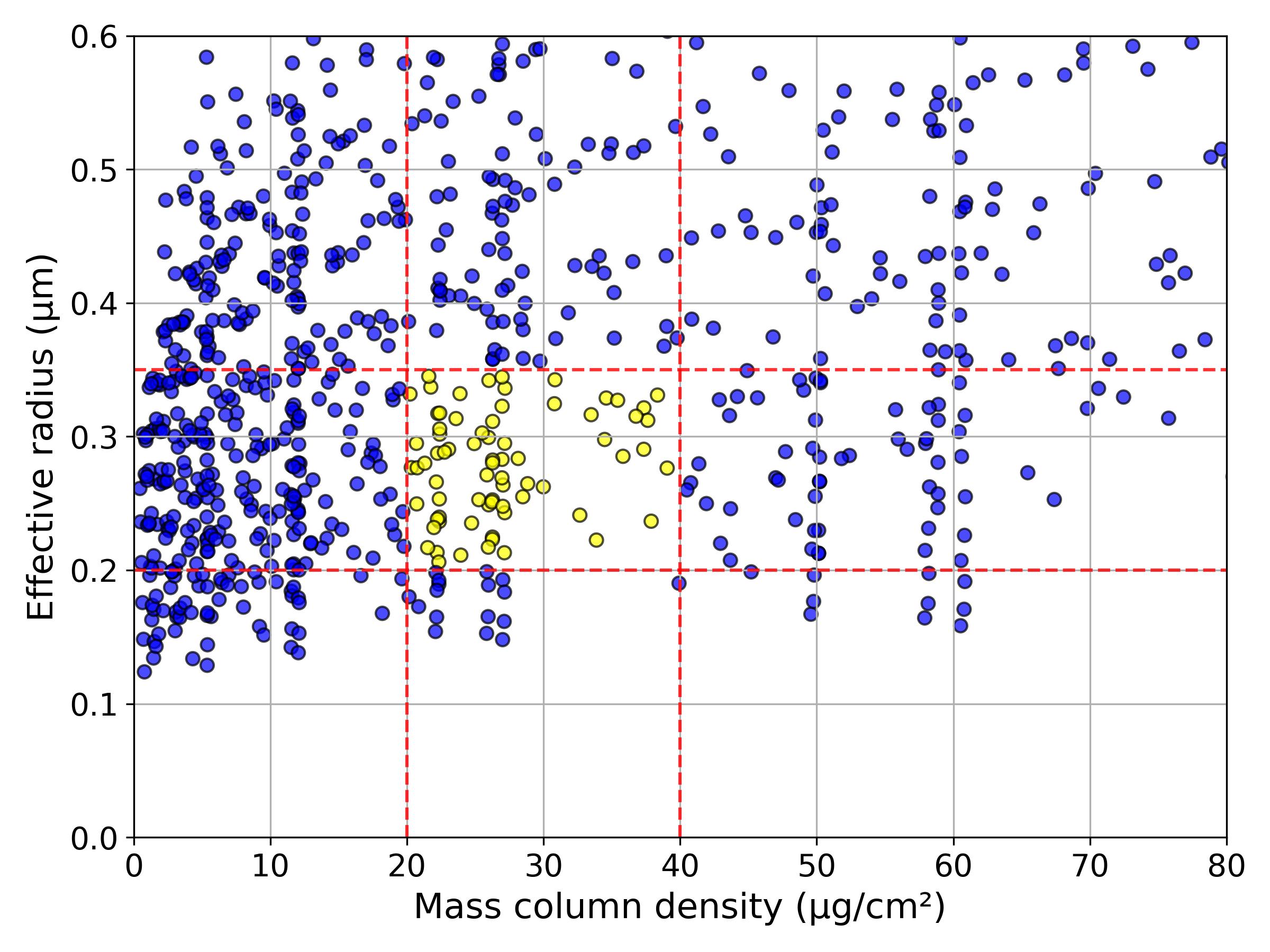}
	\caption{Output effective radius and MCD for a subset of the simulated cases. Red dashed lines indicate the target ranges used to select viable models. Yellow points correspond to simulations that simultaneously meet both criteria.  Only a portion of the full simulation set is shown for visual clarity.}
	\label{fig:grid_results}
\end{figure}

Figure~\ref{fig:grid_results} shows a representative subset of the simulated cases, shown in terms of their output effective radius and MCD. To enhance visual clarity, only a selection of the full grid is plotted, focusing on the region that spans the target ranges of $r_\mathrm{eff}=0.20$--$0.35~\mu\mathrm{m}$ and $\mathrm{MCD}=20$--$40~\mu\mathrm{g/cm^{-2}}$ defined in the previous section. A summary of the simulations that satisfy both constraints is provided in Table~\ref{tab:valid_models}, where the optical depths ($\tau$) were computed using the real and imaginary refractive indices retrieved by \citet{AnguianoArteaga2021} from Hubble Space Telescope observations using radiative transfer modelling. The convergence time reported for each case follows the criterion defined in Section \ref{model_params}.

\begin{table}[h!]
	\centering
	\caption{Subset of simulations that meet the target ranges for output effective radius and MCD.}
	\label{tab:valid_models}
	\resizebox{1.0\textwidth}{!}{%
		\begin{tabular}{ccccccccc}
			\hline
			$C_\mathrm{inj}$ & $v_\mathrm{trop}$ & $P_\mathrm{inj}$ & FWHM & $Q$ & $r_\mathrm{eff}$ & MCD & $\tau$ & $t_\mathrm{conv}$ \\
			(particles cm$^{-3}$ s$^{-1}$) & (m s$^{-1}$) & (bar) & (bar) & (electrons/$\mu$m) & ($\mu$m) & ($\mu$g/cm$^2$) & @ 900 nm & (years) \\
			\hline
			$1.6 \times 10^{-2}$ & $1.4 \times 10^{-4}$ & 0.1 & 0.09 & 30 & 0.22 & 21 & 1.40 & 14.3 \\
			$3.5 \times 10^{-2}$ & $5.0 \times 10^{-5}$ & 0.1 & 0.09 & 20 & 0.33 & 24 & 2.21 & 8.9 \\
			$3.5 \times 10^{-2}$ & $5.0 \times 10^{-5}$ & 0.1 & 0.09 & 25 & 0.28 & 28 & 2.40 & 9.8 \\
			$3.5 \times 10^{-2}$ & $5.0 \times 10^{-5}$ & 0.1 & 0.09 & 30 & 0.24 & 33 & 2.40 & 10.8 \\
			$3.5 \times 10^{-2}$ & $5.0 \times 10^{-5}$ & 0.2 & 0.18 & 25 & 0.25 & 21 & 1.54 & 10.1 \\
			$3.5 \times 10^{-2}$ & $5.0 \times 10^{-5}$ & 0.2 & 0.18 & 30 & 0.21 & 24 & 1.43 & 11.4 \\
			$3.5 \times 10^{-2}$ & $1.4 \times 10^{-4}$ & 0.1 & 0.09 & 20 & 0.34 & 31 & 2.92 & 9.8 \\
			$3.5 \times 10^{-2}$ & $1.4 \times 10^{-4}$ & 0.1 & 0.09 & 25 & 0.29 & 37 & 3.31 & 11.4 \\
			$3.5 \times 10^{-2}$ & $1.4 \times 10^{-4}$ & 0.2 & 0.18 & 20 & 0.31 & 24 & 2.12 & 10.8 \\
			$3.5 \times 10^{-2}$ & $1.4 \times 10^{-4}$ & 0.2 & 0.18 & 25 & 0.26 & 29 & 2.30 & 12.0 \\
			$3.5 \times 10^{-2}$ & $1.4 \times 10^{-4}$ & 0.2 & 0.18 & 30 & 0.22 & 34 & 2.21 & 13.6 \\
			$7.9 \times 10^{-2}$ & $5.0 \times 10^{-5}$ & 0.1 & 0.06 & 20 & 0.34 & 22 & 1.97 & 7.3 \\
			$7.9 \times 10^{-2}$ & $5.0 \times 10^{-5}$ & 0.1 & 0.06 & 25 & 0.29 & 25 & 2.14 & 7.9 \\
			$7.9 \times 10^{-2}$ & $5.0 \times 10^{-5}$ & 0.1 & 0.06 & 30 & 0.25 & 28 & 2.19 & 8.6 \\
			$7.9 \times 10^{-2}$ & $5.0 \times 10^{-5}$ & 0.2 & 0.18 & 20 & 0.33 & 35 & 3.14 & 8.2 \\
			$7.9 \times 10^{-2}$ & $1.4 \times 10^{-4}$ & 0.1 & 0.06 & 25 & 0.32 & 33 & 3.03 & 10.1 \\
			$7.9 \times 10^{-2}$ & $1.4 \times 10^{-4}$ & 0.1 & 0.06 & 30 & 0.28 & 39 & 3.27 & 11.1 \\
			$7.9 \times 10^{-2}$ & $1.4 \times 10^{-4}$ & 0.2 & 0.12 & 25 & 0.29 & 21 & 1.74 & 11.4 \\
			$7.9 \times 10^{-2}$ & $1.4 \times 10^{-4}$ & 0.2 & 0.12 & 30 & 0.25 & 25 & 1.91 & 12.7 \\
			$1.8 \times 10^{-1}$ & $5.0 \times 10^{-5}$ & 0.1 & 0.03 & 20 & 0.33 & 20 & 1.81 & 7.6 \\
			$1.8 \times 10^{-1}$ & $5.0 \times 10^{-5}$ & 0.1 & 0.03 & 25 & 0.29 & 23 & 1.94 & 8.2 \\
			$1.8 \times 10^{-1}$ & $5.0 \times 10^{-5}$ & 0.1 & 0.03 & 30 & 0.25 & 26 & 1.96 & 8.9 \\
			$1.8 \times 10^{-1}$ & $5.0 \times 10^{-5}$ & 0.2 & 0.12 & 20 & 0.35 & 22 & 1.90 & 7.3 \\
			$1.8 \times 10^{-1}$ & $5.0 \times 10^{-5}$ & 0.2 & 0.12 & 25 & 0.30 & 26 & 2.14 & 8.2 \\
			$1.8 \times 10^{-1}$ & $5.0 \times 10^{-5}$ & 0.2 & 0.12 & 30 & 0.26 & 30 & 2.28 & 9.2 \\
			$1.8 \times 10^{-1}$ & $1.4 \times 10^{-4}$ & 0.1 & 0.03 & 25 & 0.32 & 31 & 2.77 & 10.1 \\
			$1.8 \times 10^{-1}$ & $1.4 \times 10^{-4}$ & 0.1 & 0.03 & 30 & 0.29 & 36 & 3.02 & 11.1 \\
			$1.8 \times 10^{-1}$ & $1.4 \times 10^{-4}$ & 0.2 & 0.06 & 25 & 0.28 & 20 & 1.63 & 10.8 \\
			$1.8 \times 10^{-1}$ & $1.4 \times 10^{-4}$ & 0.2 & 0.06 & 30 & 0.23 & 25 & 1.72 & 12.4 \\
			$1.8 \times 10^{-1}$ & $1.4 \times 10^{-4}$ & 0.2 & 0.12 & 25 & 0.33 & 38 & 3.35 & 10.1 \\
			$4.0 \times 10^{-1}$ & $5.0 \times 10^{-5}$ & 0.2 & 0.06 & 25 & 0.29 & 23 & 1.85 & 7.9 \\
			$4.0 \times 10^{-1}$ & $5.0 \times 10^{-5}$ & 0.2 & 0.06 & 30 & 0.25 & 27 & 1.94 & 8.9 \\
			$4.0 \times 10^{-1}$ & $1.4 \times 10^{-4}$ & 0.2 & 0.06 & 25 & 0.31 & 38 & 3.21 & 9.5 \\
			\hline
		\end{tabular}%
	}
\end{table}

Across the subset of cases that simultaneously fall within the observationally inferred ranges for $r_\mathrm{eff}$ (0.2--0.35 $\mu$m) and MCD (20--40 $\mu$g/cm$^{2}$), the convergence times are never shorter than $\sim$7 Earth years. We therefore identify this $\sim$7 years as a lower bound on the characteristic build--up time of the observed chromophore layer in the GRS. Faster--growing cases cannot match the required $r_\mathrm{eff}$ and MCD simultaneously.

These viable cases occupy a restricted region of the parameter space, with injection rates in the range \(C_\mathrm{inj} \in [1.6 \times 10^{-2},\ 4.0 \times 10^{-1}]~\mathrm{cm^{-3}\,s^{-1}}\). Assuming a particle density of \(0.7~\mathrm{g\,cm^{-3}}\), the corresponding volumetric mass injection rates span \(4.7\times10^{-17}\) to \(1.2\times10^{-15}~\mathrm{kg\,m^{-3}\,s^{-1}}\);
when integrated over altitude, the associated column mass injection fluxes are \(1.1\times10^{-12}\) to \(6.9\times10^{-12}~\mathrm{kg\,m^{-2}\,s^{-1}}\). Vertical velocities associated with these cases lie in the range \(v_\mathrm{trop} \in [5.0 \times 10^{-5},\ 1.4 \times 10^{-4}]\) m s$^{-1}$, close to the lower bound estimated for the GRS tropopause by \citet{Conrath1981}. Higher values of \(C_\mathrm{inj}\) lead to shorter convergence times, as coagulation proceeds more rapidly when a greater number of particles is available. Conversely, higher values of \(v_\mathrm{trop}\) generally result in longer convergence times, given the higher fall velocity (and thus, size) needed for the particles to settle.

All simulations that satisfy the modelling constraints favour particle charge values of \(Q \geq 20~ \text{electrons}/\mu\mathrm{m}\), highlighting the importance of electrostatic repulsion in regulating coagulation and preventing excessive particle growth. As seen in Table~\ref{tab:valid_models}, longest convergence times occur for \(Q = 30~ \text{electrons}/\mu\mathrm{m}\). However, this value does not necessarily imply a slow approach to a stationary state, with \(C_\mathrm{inj}\) and \(v_\mathrm{trop}\) being more decisive. As for the injection pressure, both tested levels—0.1 and 0.2~bar—are found to produce acceptable outcomes across all three FWHM values (30\%, 60\%, and 90\% of the injection pressure), indicating that a variety of vertical injection profiles can lead to consistent simulations, provided that the injection remains within the chromophore region.

\begin{figure}[h]
	\centering
	\includegraphics[width=1.\textwidth]{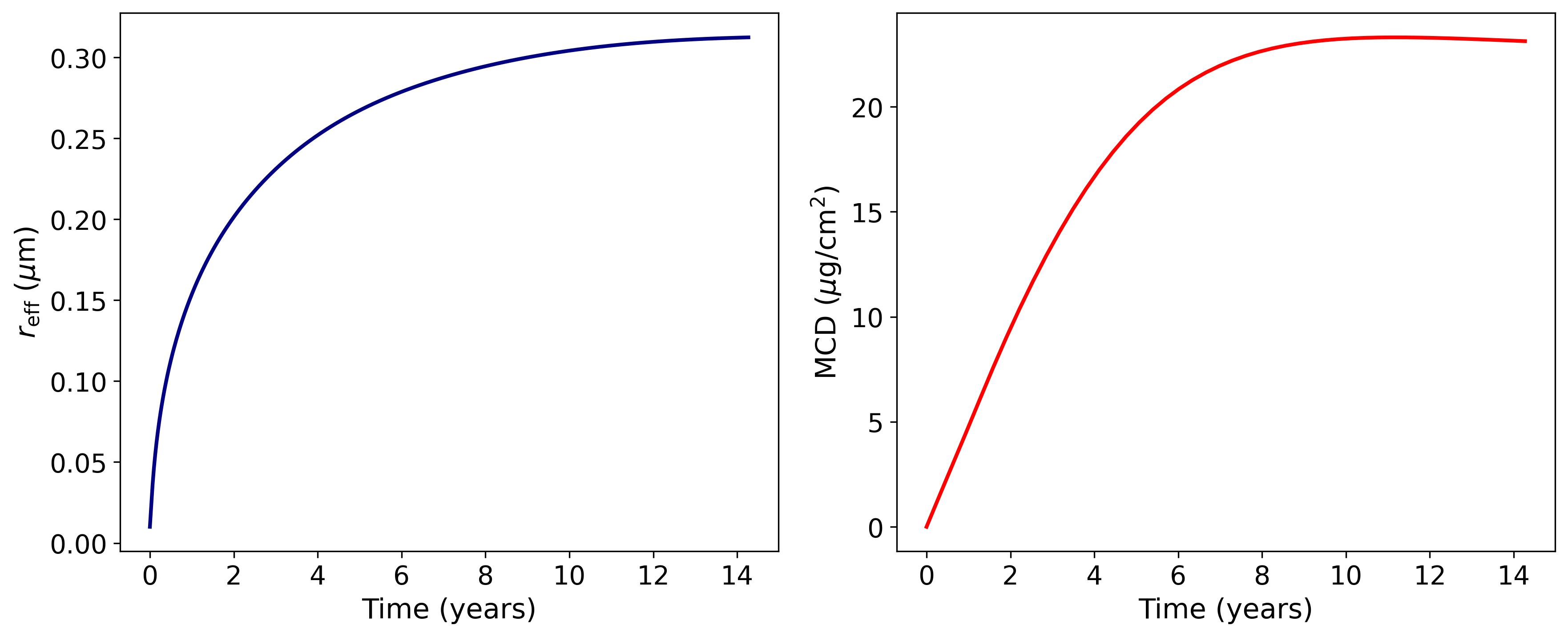}
	\caption{Time evolution of the effective radius (left) and MCD (right) for the selected reference model. Both properties asymptotically approach a stationary state after approximately a Jovian year ($\approx$ 12 Earth years).}
	\label{fig:reff_mcd_evolution}
\end{figure}

\begin{figure}[h]
	\centering
	\includegraphics[width=0.7\textwidth]{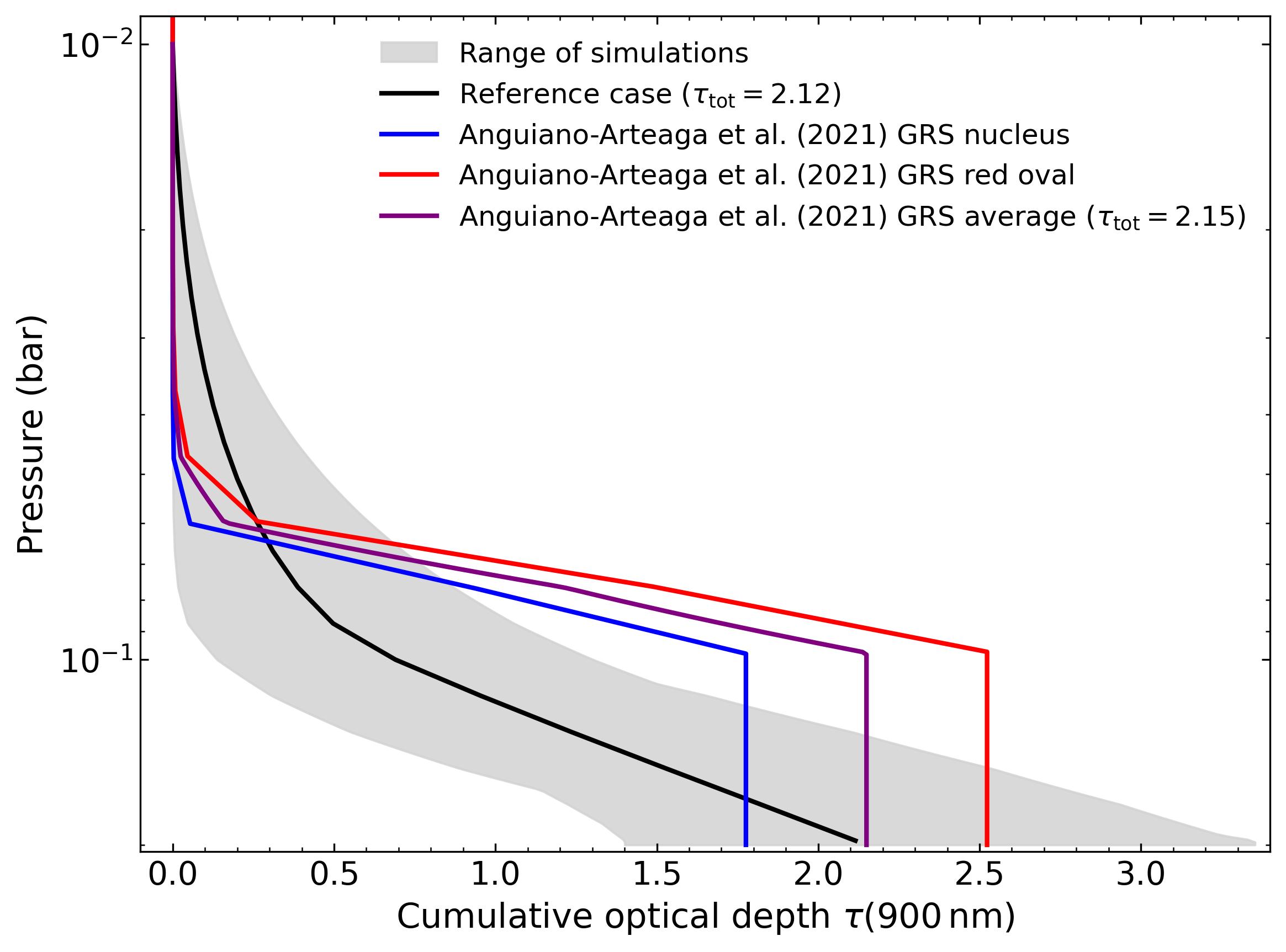}
	\caption{Cumulative optical depth at 900~nm as a function of pressure. The grey shaded region shows the range spanned by the successful microphysical simulations listed in Table~\ref{tab:valid_models}, and the solid black line highlights the reference case. The blue and red curves show the chromophore profiles retrieved from radiative transfer modelling by \citet{AnguianoArteaga2021} for the GRS nucleus and its surrounding red oval, respectively, while the purple curve shows the average of these two profiles. In those retrievals, the base of the chromophore layer lies near \(P = 0.1\)~bar, so the cumulative optical depth remains approximately constant at higher pressures.}
	\label{fig:tau_profiles}
\end{figure}

The P--T profile shown in Figure~\ref{fig:ptprofile}, as explained in Section~\ref{atm_profiles}, was constructed by averaging over a 3×3 pixel region within the GRS's reddish area. Given the spatial resolution of the temperature maps in \citet{Fletcher2020}, this region encompasses both the GRS nucleus and its reddish surroundings. To select a representative simulation, we adopted as benchmarks the averages reported by \citet{AnguianoArteaga2021} for those two regions: \(r_\mathrm{eff}=0.31~\mu\mathrm{m}\), \(\mathrm{MCD}=25~\mu\mathrm{g\,cm^{-2}}\), and \(\tau(900\,\mathrm{nm})=2.15\). The simulation that most closely matches these targets corresponds to an injection rate of \(C_\mathrm{inj} = 3.5 \times 10^{-2}~\mathrm{cm^{-3}\,s^{-1}}\), a vertical tropospheric velocity of \(v_\mathrm{trop} = 1.4 \times 10^{-4}\) m s$^{-1}$, an injection pressure of 0.2~bar, a FWHM of 0.18~bar, a particle charge of \(Q = 20~\mathrm{electrons}/\mu\mathrm{m}\) and a total cumulative optical depth of \(\tau(900\,\mathrm{nm})=2.12\). The temporal evolution of the effective radius and MCD for this reference case is shown in Figure~\ref{fig:reff_mcd_evolution}. Both quantities progressively increase and converge toward steady values after approximately 11 years of simulated time, indicating that the system reaches equilibrium on that timescale. The cumulative optical depths spanned by the successful simulations in Table~\ref{tab:valid_models}, together with the corresponding curve of the reference case and the analogous chromophore profiles retrieved from radiative transfer modelling by \citet{AnguianoArteaga2021} for the GRS nucleus and its surrounding reddish interior, are shown in Figure~\ref{fig:tau_profiles}.

\begin{figure}[h]
	\centering
	\includegraphics[width=1.\textwidth]{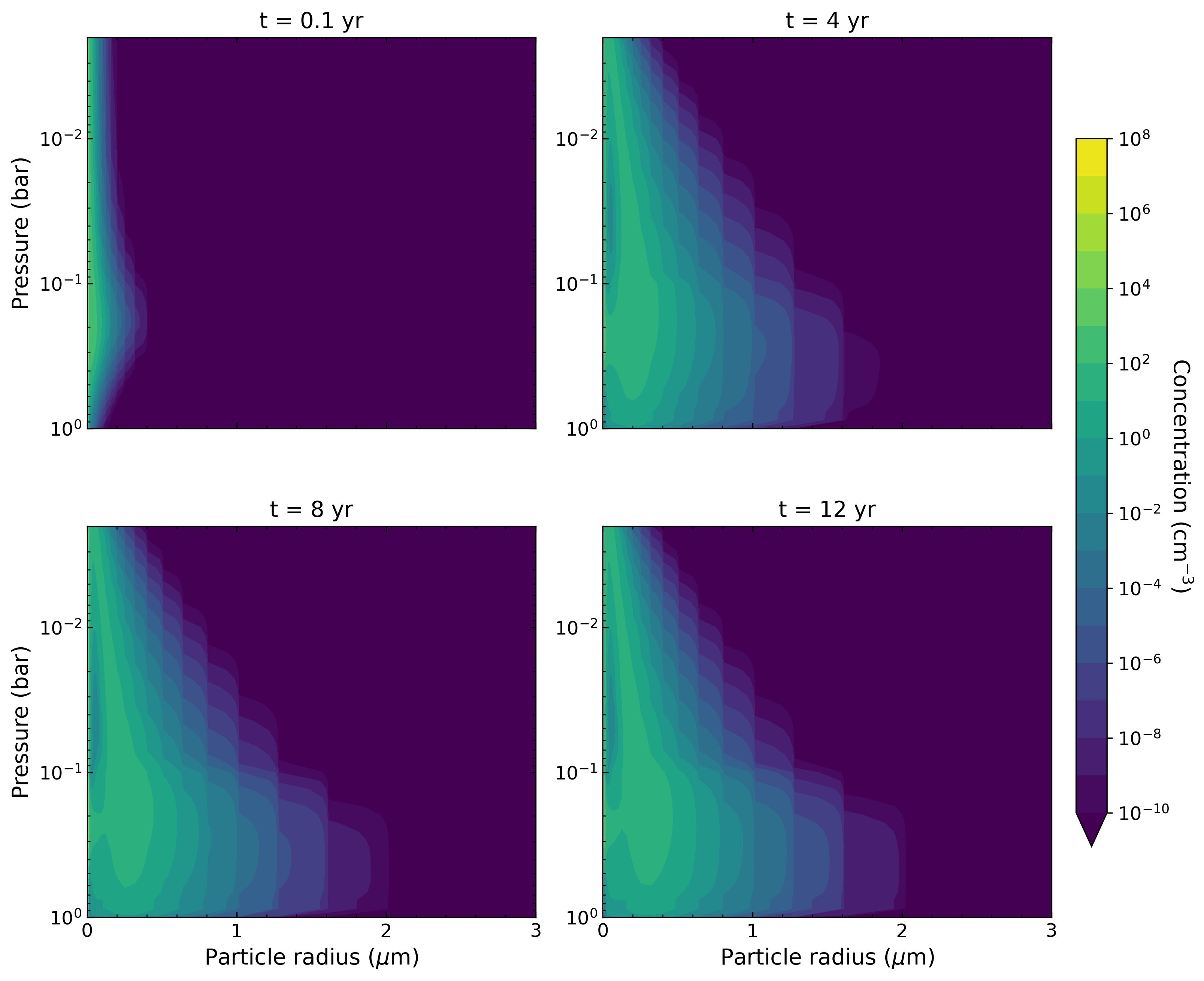}
	\caption{Temporal evolution of the particle size distribution and vertical concentration profile at four selected times (0.1, 4, 8, and 12 years). Each panel shows the concentration as a function of particle radius and pressure. This figure corresponds to the reference simulation with $C_\mathrm{inj}=3.5\times10^{-2}~\mathrm{particles\,cm^{-3}\,s^{-1}}$, $v_\mathrm{trop}=1.4\times10^{-4}~\mathrm{m\,s^{-1}}$, $P_\mathrm{inj}=0.2~\mathrm{bar}$, $\mathrm{FWHM}=0.18~\mathrm{bar}$, and $Q=20~\mathrm{electrons/\mu m}$.} 
	\label{fig:concentration_evolution}
\end{figure}

The vertical and size distribution of particles at different stages of the reference simulation is presented in Figure~\ref{fig:concentration_evolution}, where the growth by coagulation and sedimentation of larger particles can be observed, along with upward diffusive transport. At early times, strong diffusive transport from the injection region maintains relatively high concentrations of the smallest particles above 0.1~bar; at later times, upward--transported larger particles efficiently scavenge the smallest size bin via coagulation, leading to the reduced concentrations of the smallest particles between 0.1 and 0.01~bar in Figure~\ref{fig:concentration_evolution}, while at lower pressures coagulation becomes less effective due to the scarcity of large particles. This interplay of microphysical processes leads to a vertically extended haze structure with a broad particle size range, stabilizing over time.  Some of the characteristic features of this type of simulation, together with its physical plausibility and relevance to previous radiative transfer modelling results, are discussed in the following section.

\clearpage
\section{Discussion}
\label{discussion}

\subsection{Origin of the GRS chromophore}
\label{chrom_origin}
\begin{figure}[h]
	\centering
	\includegraphics[width=0.95\textwidth]{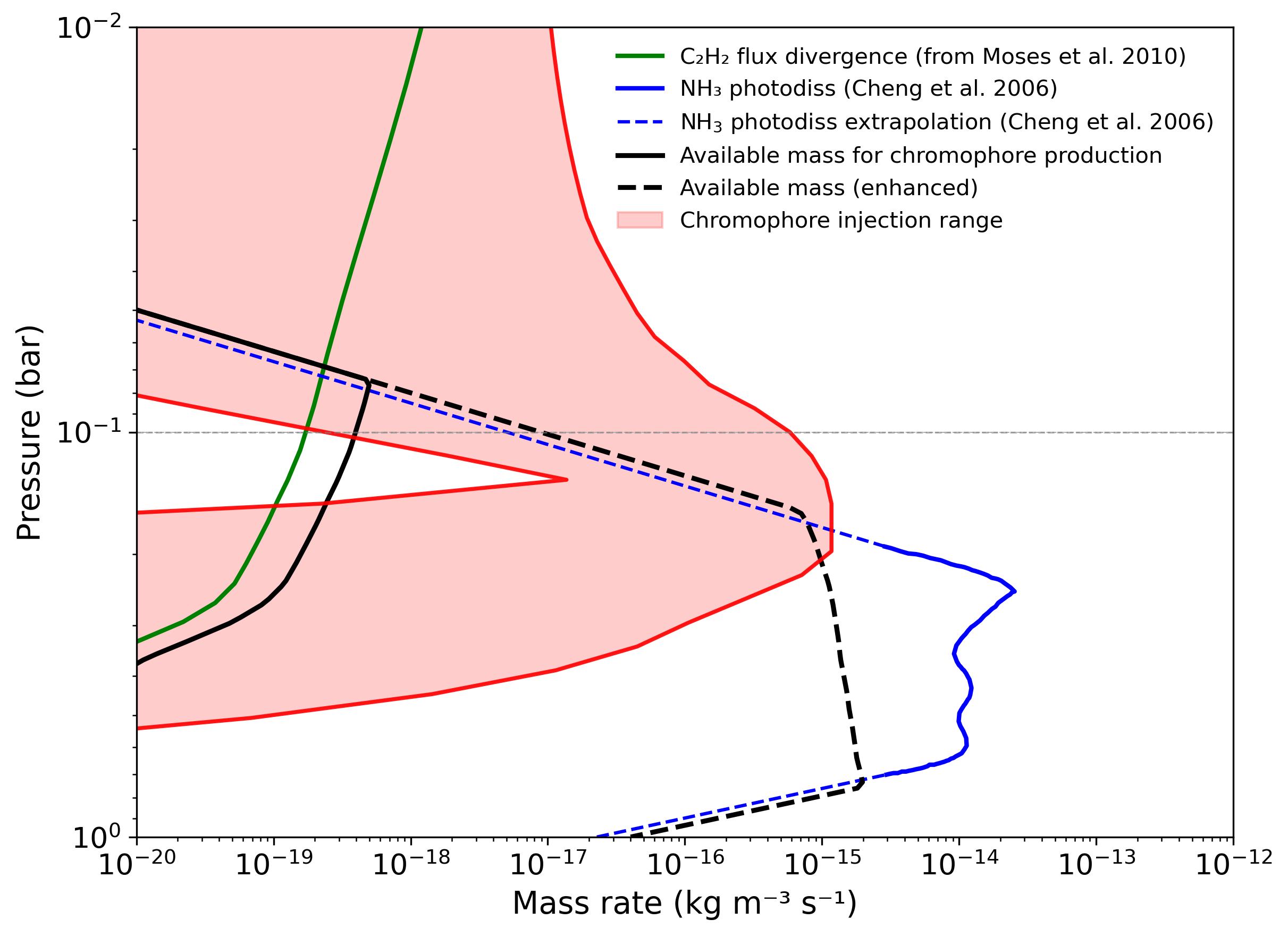}
	\caption{Comparison between retrieved chromophore mass production rates and the mass availability from parent gases. The green line shows the C$_2$H$_2$ flux divergence from \citet{Moses2010}, while the blue line corresponds to the NH$_3$ photodissociation rate from \citet{Cheng2006}, dashed segments showing extrapolated regions. The solid black curve is the nominal available mass for chromophore production; and the black dashed curve is the lightning--enhanced available mass. The shaded red region indicates the range of chromophore injection profiles consistent with our microphysical simulations.}
	\label{fig:chromophore_injection}
\end{figure}

In Figure~\ref{fig:chromophore_injection}, we compare the retrieved chromophore mass production profiles with the availability of parent gases. We compute the C$_2$H$_2$ vertical flux divergence (i.e., the local rate of supply per unit volume) from the flux in the supplementary material of \citet{Moses2010}, and take the NH$_3$ photodissociation rate from \citet{Cheng2006}. In constructing the available mass curve below the tropopause, where NH$_3$ remains abundant, we adopt the working assumption that two NH$_3$ molecules are available per C$_2$H$_2$ molecule, motivated by the two--N--bearing composition of the Carlson--type chromophores. Above the tropopause, however, NH$_3$ scarcity may limit production. To preserve the two--NH$_3$ per unit interpretation without overspending NH$_3$, we enforce a limiting--reagent condition at each pressure: we compare the molecular supply of C$_2$H$_2$ with half the molecular rate of NH$_3$ and adopt the lower value. The resulting limiting molecular rate is then converted to mass by multiplying by the combined mass of one C$_2$H$_2$ plus two NH$_3$, yielding an availability curve that is valid both below and above the tropopause. In addition, we derive an “enhanced” availability by computing an enhanced C$_2$H$_2$ flux $\Phi_{\rm enh}$ from the nominal flux $\Phi_{\rm nom}$ \citep{Moses2010}, the retrieved mole fraction ratio $R(P)=f_{\rm enh}/f_{\rm nom}$, and a diffusive--transport adjustment, $\Phi_{\rm enh}=R \cdot \Phi_{\rm nom}-K_{zz} \cdot N \cdot f_{\rm nom} \cdot \mathrm{d}R/\mathrm{d}z$, where $f_{\rm nom}$ and $f_{\rm enh}$ are the nominal and lightning--enhanced C$_2$H$_2$ mole fractions reported by \citet{Moses2010}, respectively. We then take the vertical divergence of  $\Phi_{\rm enh}$ under the same limiting--reagent assumption.

As can be seen from Figure~\ref{fig:chromophore_injection}, the successful simulations exceed the nominal available mass for chromophore formation (i.e., without enhancement), with C$_2$H$_2$ supply as the principal limiting factor in the troposphere. Furthermore, only a fraction of the available mass would be converted into chromophore material, depending on the chemical efficiency of the formation pathways. Using this nominal available mass profile as the input injection profile in our microphysical model, we find that the system reaches a steady state after approximately 10--25~years (depending on the assumed $v_{\rm trop}$ and $Q$), with a resulting chromophore MCD of about 0.1~$\mu$g/cm$^{2}$, well below the observational target range. When using an enhanced injection based on the lightning--enhanced C$_2$H$_2$ case, the model outputs approach the observational target range: our closest overall match occurs for $v_{\rm trop}=5\times10^{-5}$ m s$^{-1}$ and $Q=25$ electrons/$\mu$m, yielding $r_{\rm eff}=0.35~\mu$m, MCD $= 41~\mu$g/cm$^2$ and a convergence time of approximately 6 years. Simulations that fall within the desired $r_{\rm eff}$ and MCD ranges are obtained for chemical efficiencies in the range of 20--95\% in $\sim$7--12 years. However, the plausibility of a C$_2$H$_2$ enhancement by lightning remains debated \citep{Moses2010,Baines2019}. Comparatively large tropospheric C$_2$H$_2$ mole fractions, such as those reported by \citet{Betremieux2003}, could in principle meet the required production rates, but \citet{Moses2010} provide strong arguments against such high values, noting that they would generate absorption wings around the mid--infrared C$_2$H$_2$ emission lines that are not observed. Predicted abundances from different photochemical models often differ by more than an order of magnitude \citep{Knizek2026}. Consequently, establishing the C$_2$H$_2$ budget available for chromophore formation in the GRS requires observational constraints on the upper--tropospheric C$_2$H$_2$ abundance within the GRS itself. Existing observations may already be suitable for this purpose: Voyager/IRIS and Cassini/CIRS spectra have been employed to retrieve zonally--averaged tropospheric C$_2$H$_2$ \citep{Nixon2010}, while stratospheric C$_2$H$_2$ abundances have been retrieved from both space--based JWST/MIRI observations \citep{RodriguezOvalle2024} and ground--based mid--IR measurements with IRTF/TEXES \citep{Fletcher2016,Melin2018}. In this context, both TEXES and JWST/MIRI offer higher spectral resolution than Voyager/IRIS or Cassini/CIRS in the $\sim$14~$\mu$m region, where the emission lines normally used to retrieve C$_2$H$_2$ abundances are located. Although constraining the tropospheric C$_2$H$_2$ abundance in the GRS is key to determining the budget available for the Carlson chromophore, the available constraints and most current photochemical models for Jupiter tend to indicate a shortfall in upper--tropospheric C$_2$H$_2$ to sustain the mass injection rate required for a haze layer composed exclusively of the Carlson chromophore.

As proposed by \citet{Carlson2016}, an enhanced upward flux of NH$_3$ into the high troposphere could increase the material available for chromophore formation. Enhanced injection at higher altitudes is compatible with our range of injection profiles in Fig.~\ref{fig:chromophore_injection}. However, current observations show no persistent, large--scale enhancement of NH$_3$ in the upper troposphere of the Great Red Spot, as shown by recent analyses of JWST/MIRI \citep{Harkett2024} and Juno/JIRAM data \citep{Grassi2021}.

Our study indicates that C$_2$H$_2$ is the limiting reagent for chromophore formation at 0.1--0.2 bar. Various photochemical models of Jupiter predict a mole fraction of $\sim10^{-9}$ near the tropopause \citep{Moses2005,Moses2010,Hue2018}. We explore if other hydrocarbons, however, could also play a role. Models also predict mole fractions of C$_2$H$_6$ that exceed that of  C$_2$H$_2$, yet our analysis of flux divergences from \citet{Moses2010} shows that their local volumetric supply between 0.3 bar and 1 mbar are similar. C$_2$H$_6$ is known to be less reactive than C$_2$H$_2$, and its photolyzation also requires higher--energy photons, which penetrate less deeply. For the next most abundant hydrocarbon, C$_2$H$_4$, the flux divergence is two orders of magnitude lower. We therefore infer that the Carlson chromophore is unlikely to be sourced primarily from C$_2$H$_4$ or C$_2$H$_6$ at these levels. Secondary contributions, however, cannot be ruled out. Robust quantification would require a GRS--tailored photochemical model that considers all plausible pathways and realistic, potentially enhanced, vertical transport.

As the C$_2$H$_2$+NH$_3$ supply appears insufficient to account for the full retrieved column mass of the chromophore haze, the contribution of additional aerosol constituents to the total mass of the layer cannot be ruled out. One possibility is an influx of aerosols formed at higher stratospheric altitudes following CH$_4$ photodissociation and subsequently transported downward. Alternatively, additional species could be supplied from deeper levels via vertical diffusion and potential upward advection. In this context, PH$_3$ is a natural candidate given its relative abundance in Jupiter’s upper troposphere \citep{SanchezLavega2011}. Consistent with this, \citet{Harkett2024} report an enhancement of PH$_3$ above the GRS and a correlation with increased upper--tropospheric aerosol opacity, suggesting that upper--tropospheric PH$_3$ could contribute to the haze bulk mass. PH$_3$ photochemistry has been proposed as a potential source of red--coloured products in Jupiter’s atmosphere \citep{Atreya1986}. However, laboratory simulation experiments by \citet{VeraRuizRowland1978} indicate that key PH$_3$ photolysis intermediates (PH, PH$_2$, P$_2$) are efficiently scavenged by C$_2$H$_2$ and C$_2$H$_4$, suppressing the formation of red--phosphorus solids and leading to the formation of a white, powder--like material. Such PH$_3$--hydrocarbon products could still contribute to the aerosol total mass. Assessing this possibility requires improved laboratory constraints on the optical properties of the Carlson chromophore. \citet{AnguianoArteaga2021} retrieved imaginary refractive indices with a spectral slope consistent with those reported by \cite{Carlson2016} but with a lower magnitude, which they suggested could reflect mixing of the chromophore with material that does not contribute appreciably to the short--wavelength absorption. However, the imaginary refractive indices provided by \citet{Carlson2016} corresponded to a specific irradiation time for their laboratory sample, chosen primarily for instrumental considerations. Establishing these indices under Jupiter--representative conditions would therefore enable a more robust assessment of mixing between the chromophore and additional aerosol constituents.

\subsection{Particle size and vertical distributions}
\label{RT_discuss}
\begin{figure}[t]
	\centering
	\includegraphics[width=\linewidth]{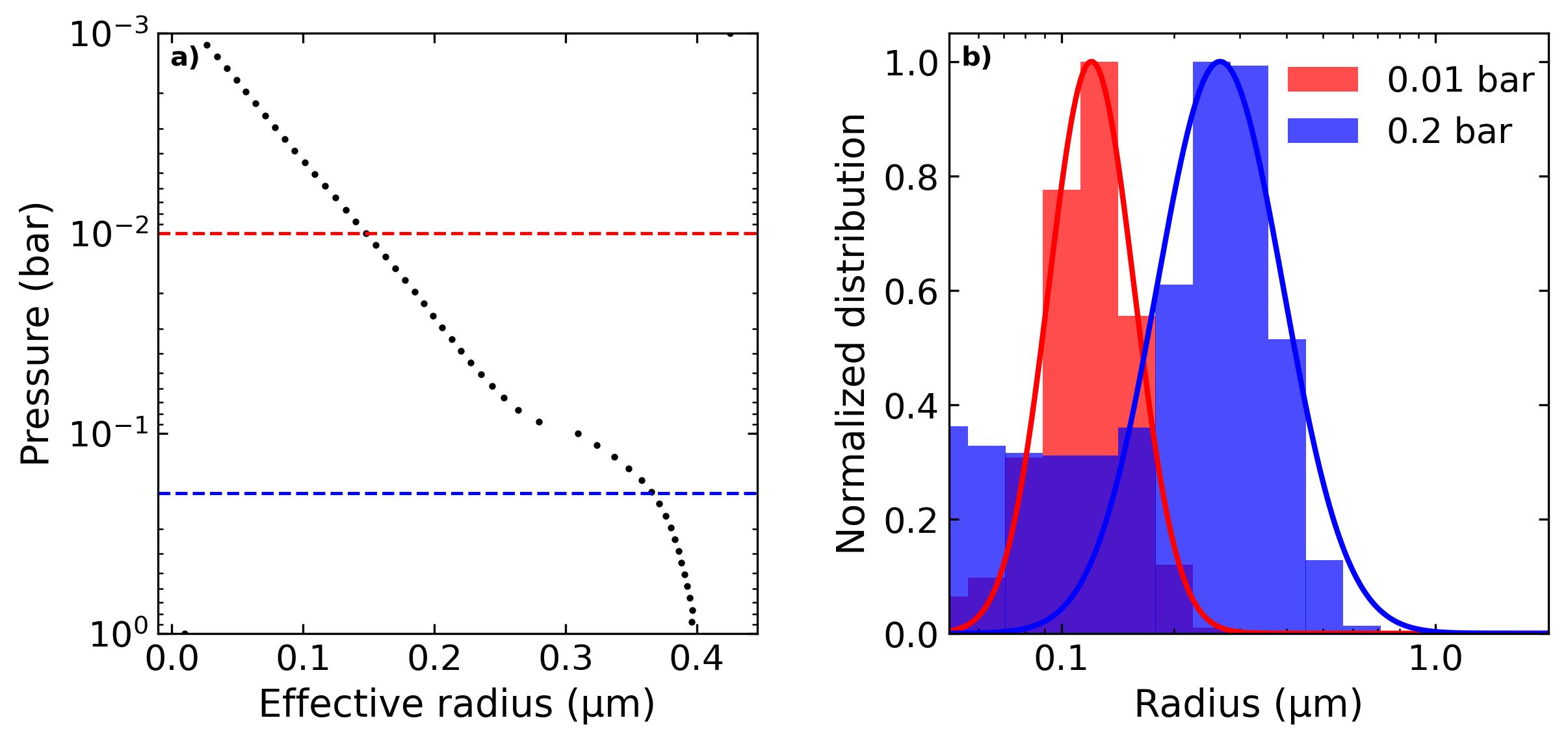}
	\caption{For the reference simulation:
		(a) Effective radius $r_{\mathrm{eff}}$ versus pressure; dashed lines mark the two selected pressure levels.
		(b) Normalized histograms (peak--normalized to 1) with a size cutoff $r \ge 0.05\,\mu$m and the corresponding log--normal fit (solid line).}
	\label{fig:size-dist}
\end{figure}

In our simulations, the effective particle radius increases monotonically with pressure, as shown in Figure~\ref{fig:size-dist}. A distinct change in curvature occurs near the tropopause as a consequence of the structure of the eddy diffusion coefficient profile \(K_{zz}\) (Figure~\ref{fig:Kprofile}). This change is not caused by the tropospheric vertical velocity, as it persists even when $v_{trop}$ is set to zero, albeit yielding slightly smaller tropospheric particles. In our setup, \(K_{zz}\) reaches a minimum near the tropopause and then increases toward higher pressures. Since the residence time for collisional growth scales as \(\tau_{\mathrm{mix}}\!\sim\!H^{2}/K_{zz}\) (with $H$ a scale height), particles spend longer near this minimum and grow more efficiently there; as \(K_{zz}\) rises with depth, \(\tau_{\mathrm{mix}}\) shortens and transport limits further growth, yielding the near--asymptotic behavior of \(r_{\mathrm{eff}}\) at depth. With a constant--log--slope \(K_{zz}\) (decreasing with pressure), \(r_{\mathrm{eff}}\) grows more steeply below the tropopause and approximately linearly with \(\log P\) and the near--asymptotic deep behavior no longer appears. 

Figure~\ref{fig:size-dist} also shows particle size distributions at \(\,0.01\,\)bar and \(\,0.2\,\)bar. When the full size range is considered, the distributions peak at the minimum radius, reflecting the injected seed particles. However, these small particles contribute weakly to light extinction: more than 99\% of total optical depth is due to particles with radii above 0.1 $\mu$m. Motivated by this result, the right plot (panel~b) excludes small particles and overlays fits to a log--normal distribution \citep{HansenTravis1974}. At \(0.01\)~bar the particle size distribution is well described by a log--normal distribution, whereas at \(0.2\)~bar the distribution exhibits a low--radius wing that a single log--normal cannot reproduce. The low--radius wing is consistent with an interplay between coagulation growth and vertical exchange around the injection peak at \(0.2\)~bar. When \(K_{zz}\) is prescribed with a constant slope in \(\log P\) (yielding smaller \(K_{zz}\) at \(0.2\)~bar), vertical exchange is reduced, increasing residence times and enhancing coagulation near the injection peak, so the low--radius wing is suppressed. The wing persists for \(v_{\rm trop}=0\), indicating that it is not caused by a sedimentation bottleneck associated with the imposed upwelling. Despite the low--radius wing, the conventional log--normal assumption remains appropriate for radiative transfer parameterisations, as extinction is dominated by larger particles: in our simulations, particles with \(r>0.2~\mu\mathrm{m}\) account for over 97\% of the column optical depth.

A widely used parameterisation in radiative transfer calculations is the Crème--Brûlée layout, in which the chromophore is located in a thin layer (extending from 0.20 to 0.18 bar) on top of the tropospheric NH$_3$ cloud \citep{Sromovsky2017, Baines2019}. This scheme is not supported by our calculations. We tested chromophore configurations with very narrow injections near 0.20 bar and found that the layers broaden rapidly, as diffusion quickly smooths steep vertical gradients and spreads particles vertically. Consistent with this behavior, the optical depth in the models of Table~\ref{tab:valid_models} is not strongly concentrated in a thin sheet: only about 15\% of the column optical depth arises from 0.20--0.18 bar, whereas typically more than 75\% originates between 0.20 and 0.10 bar. Additional mechanisms that counteract the diffusive flux would be required to confine the particle concentration in narrow layers.  

\section{Conclusions}
\label{conclusions}
We have developed a one--dimensional microphysical model, based on the computational analogs of \citet{Toon1988}, that includes sedimentation, eddy diffusion and coagulation. Using physically plausible parameters and realistic timescales, we have reproduced the effective radii and mass column densities inferred for the \citet{Carlson2016} chromophore in the Great Red Spot by recent radiative transfer studies \citep{Baines2019, Braude2020, AnguianoArteaga2021}. After parameterising the net gas--to--particle conversion via the prescribed injection term, no explicit condensational--growth process is required to match these values. Our main conclusions are as follows:

\begin{itemize}
	\item \textbf{Formation timescale.}
	For parameter combinations that reproduce the previously reported effective radii ($\sim$0.2--0.35\,$\mu$m) and mass column densities (20--40\,$\mu$g/cm$^{2}$), the chromophore haze in the GRS requires at least 7~Earth years of continuous injection to approach steady state. This sets a lower bound on the characteristic formation time of the haze and provides the first quantitative constraint on the build--up timescale of the GRS chromophore. 	
	
	\item \textbf{Chromophore production rate.} The minimum column--integrated injection rate required by our successful models is  \( \sim 10^{-12}~\mathrm{kg\,m^{-2}\,s^{-1}}\). The needed rates exceed the mass available from parent gases implied by photochemical models—C$_2$H$_2$ flux divergences from \citet{Moses2010} and NH$_3$ photodissociation from \citet{Cheng2006}. In the troposphere C$_2$H$_2$ is the limiting reagent, whereas above the tropopause photolyzed NH$_3$ limits production. This implies that higher C$_2$H$_2$ tropospheric abundances and associated fluxes could meet the required budget. Notably, existing photochemical models are not GRS--specific, and constraining the upper--tropospheric C$_2$H$_2$ abundance at $\sim$0.1--0.2~bar within the GRS itself is therefore key to establishing the C$_2$H$_2$ budget available for Carlson--chromophore production. Nevertheless, the available constraints and most current photochemical models for Jupiter suggest that upper--tropospheric C$_2$H$_2$ is insufficient to sustain the mass injection rate required for a haze layer composed exclusively of the Carlson chromophore. The possibility of other aerosol constituents contributing to the bulk mass of the chromophore haze layer cannot be discarded, so that the Carlson chromophore need not account for the entire retrieved column mass. As already pointed out by \citet{Baines2019}, lightning--driven C$_2$H$_2$ production \citep{BarNun1985,Podolak1988} could in principle raise tropospheric C$_2$H$_2$ to levels sufficient to support the required chromophore production rates, although the plausibility of this mechanism remains debated. The JUICE mission is expected to deliver a statistical view of Jovian lightning activity across different atmospheric regions \citep{Fletcher2023}, offering a way to evaluate this possibility.	
			
	\item \textbf{Size--distribution parameterisation.} Above the optically relevant cutoff ($r \gtrsim 0.1~\mu$m), the simulated particle size distributions are well represented by a single log--normal at the pressure levels that dominate extinction. A log--normal parameterisation is therefore adequate for radiative transfer retrievals of the Carlson type chromophore.
	
	\item \textbf{Vertical layout.} A thin, Crème--Brûlée--type sheet confined to $\sim$0.20--0.18~bar is not favoured in our model: eddy diffusion rapidly broadens any narrowly injected layer, and the steady--state optical depth is vertically distributed rather than concentrated in a thin slab. Additional processes that counteract the diffusive flux would be required to maintain a persistent, sharply confined chromophore layer.
	
\end{itemize}

We identify three priorities to solidify the Carlson‐type chromophore scenario. (i) Constrain the chromophore--production budget: measurements of C$_2$H$_2$ mixing ratios at $\sim$0.2 bar and associated fluxes in the GRS are needed to better quantify the available C$_2$H$_2$ for chromophore formation. A GRS--specific photochemical model, including accurate vertical transport and all plausible pathways to the Carlson--family compounds, would enable quantification of chromophore--production efficiencies under realistic conditions. Stratospheric NH$_3$ measurements would help discern whether a portion of the chromophore is produced above the tropopause. As additional tracers, enhanced HCN and CH$_3$CN mixing ratios would be consistent with strengthened N--bearing photochemistry \citep{Moses2010,Baines2019}. (ii) Radiative transfer modelling: our analysis builds on radiative transfer retrievals of the GRS, an intrinsically degenerate problem that affects estimates of the chromophore vertical location, size distribution, and abundance. Degeneracies can be mitigated by modelling spectra over an extended wavelength range, ideally from UV to thermal IR, at high spectral resolution and for multiple observing geometries. As an additional means of identifying the most probable parameter combinations defining vertical haze structure, Bayesian inference techniques have already been employed for Solar System giant planets (e.g., Uranus; \citealp{deKleer2015}) and for exoplanets \citep{RoyPerez2025}. Notably, the widely used NEMESIS radiative transfer suite \citep{Irwin2008} now includes a Bayesian enabled implementation \citep{Alday2025}. Such techniques, applied to observations with the characteristics described above, could also be used to test the vertical profiles of aerosol properties inferred in this work. As a concrete example, Cassini/VIMS Jupiter observations provide coverage from the visible to thermal IR together with a wide range of viewing geometries (including intermediate- and high-phase angles not accessible from Earth), while complementary UV constraints could be added from Cassini/ISS. (iii) Laboratory constraints: precise spectral modelling requires robust optical properties for Carlson--type materials, specifically tightly constrained complex refractive indices determined over the UV--IR range under Jupiter--like conditions. In addition, because the Carlson family is broad, the subset of compounds whose spectra are compatible with Jovian absorption should be identified and prioritised.

\section*{CRediT authorship contribution statement}

\noindent\textbf{Asier Anguiano-Arteaga:} Conceptualization, Methodology, Software, Validation, Formal analysis, Investigation, Data curation, Visualization, Writing -- original draft, Writing -- review \& editing.\\
\noindent\textbf{Santiago Pérez-Hoyos:} Conceptualization, Supervision, Project administration, Funding acquisition, Methodology, Writing -- review \& editing.\\
\noindent\textbf{Agustín Sánchez-Lavega:} Supervision, Project administration, Funding acquisition, Writing -- review \& editing.\\
\noindent\textbf{Patrick G.J. Irwin:} Resources, Supervision, Writing -- review \& editing.

\section*{Declaration of competing interest}

The authors declare that they have no known competing financial interests or personal relationships that could have appeared to influence the work reported in this paper.

\section*{Data availability}

Data will be made available upon request.

\section*{Acknowledgments}
We are very thankful to D.~Toledo for insightful advice on microphysical modelling, including guidance on the treatment of haze formation and on the assessment of steady-state convergence.

This work was supported by the Basque Government (Grupos de Investigación, IT1742-22), Elkartek KK-2025/00106 and by Grant PID2023‐-149055NB-‐C31 funded by MICIU/AEI/10.13039/501100011033 and by FEDER, UE

A.~Anguiano-Arteaga was supported by the \textit{Programa de Perfeccionamiento de Personal Investigador Doctor 2024--2027} of the Basque Government.

\clearpage
\appendix

\renewcommand{\theequation}{A.\arabic{equation}}
\renewcommand{\thefigure}{A.\arabic{figure}}
\renewcommand{\thetable}{A.\arabic{table}}

\renewcommand{\theHequation}{A.\arabic{equation}}
\renewcommand{\theHfigure}{A.\arabic{figure}}
\renewcommand{\theHtable}{A.\arabic{table}}

\renewcommand{\figurename}{Figure}
\renewcommand{\tablename}{Table}

\setcounter{equation}{0}
\setcounter{figure}{0}
\setcounter{table}{0}
\section{Validation of Computational Algorithms}
\label{app1}

Following \citet{Toon1988}, Eq.~\ref{eq:continuity} can be split into transport and coagulation sub-equations (here we assume no particle injection, i.e., $C_{\mathrm{inj}}(z,r)=0$):

\setlength{\jot}{10pt} 

\begin{align}
	\frac{\partial C(z,r)}{\partial t} 
	&-  \frac{\partial \left[ \rho \cdot K_{zz}(z) \right] }{\partial z} \cdot \frac{\partial}{\partial z} \left[ \frac{C(z,r)}{\rho} \right]
	+ \frac{\partial \left[ W(z,r) \cdot C(z,r) \right]}{\partial z} = 0 \label{eq:transport_term} \\
	\frac{\partial C(z,r)}{\partial t} 
	&= P_{\text{coag}}(z,r) - L_{\text{coag}}(z,r) \label{eq:coag_term}
\end{align}

The first equation describes vertical transport (Eq.~\ref{eq:transport_term}), while the second governs the coagulation processes responsible for the production and loss of particles of a given size at a given altitude (Eq.~\ref{eq:coag_term}). As discussed by \citet{Toon1988}, the vertical transport equation can be formulated in such a way that it allows for either an explicit or implicit numerical solution. In our implementation, we generally prefer the implicit approach, as it prevents negative concentrations and ensures numerical stability even when using relatively large time steps.

To validate our implementation of vertical transport, we compare its behavior against known analytical solutions, focusing on cases of pure diffusion, diffusion with advection, and pure advection. 

In the case of pure diffusion, we compare our numerical results with the analytical solution to Eq.~\ref{eq:transport_term}, as presented by \citet{Toon1988}:

\begin{align}
	\begin{split}
		C(t,z) = \ &C_{0}/(2\sqrt{\pi K_{zz}t})\left\{\text{exp}[-(z-z_{0})^2/(4K_{zz}t)]\right\} 
		\\
		&+ \text{exp}[-(z+z_{0})^2/(4K_{zz}t)]  	
		\\
		&\cdot \text{exp}[-W(z-z_{0})/(2K_{zz})-W^{2}t/(4K_{zz})]	
		\\
		&+ C_{0}W\text{exp}(-Wz/K_{zz})/(K_{zz}\sqrt{\pi})
		\\
		&\cdot \int_{(z+z_{0}-Wt)/(2\sqrt{K_{zz}t})}^{\infty} e^{-\gamma^{2}}d\gamma
	\end{split}
	\label{eq:analytical_diff}
\end{align}

where \( C_0 \) is the initial concentration, modeled as a Dirac delta function centered at altitude \( z_0 \). Note that \( \gamma \) is simply a variable of integration. We also corrected a typographical error in the equation as printed in \citet{Toon1988}, where the coefficient should read \( 2\sqrt{\pi K_{zz} t} \) instead of \( 2\pi K_{zz} t \), following the original form given in \citet{HidyBrock2013}, in which also appears an additional sign typo in the 1970 edition of the original source (where a `+' sign mistakenly appears in the lower limit of the integral instead of a `-').

\begin{figure}[h]
	\centering
	\includegraphics[width=0.75\textwidth]{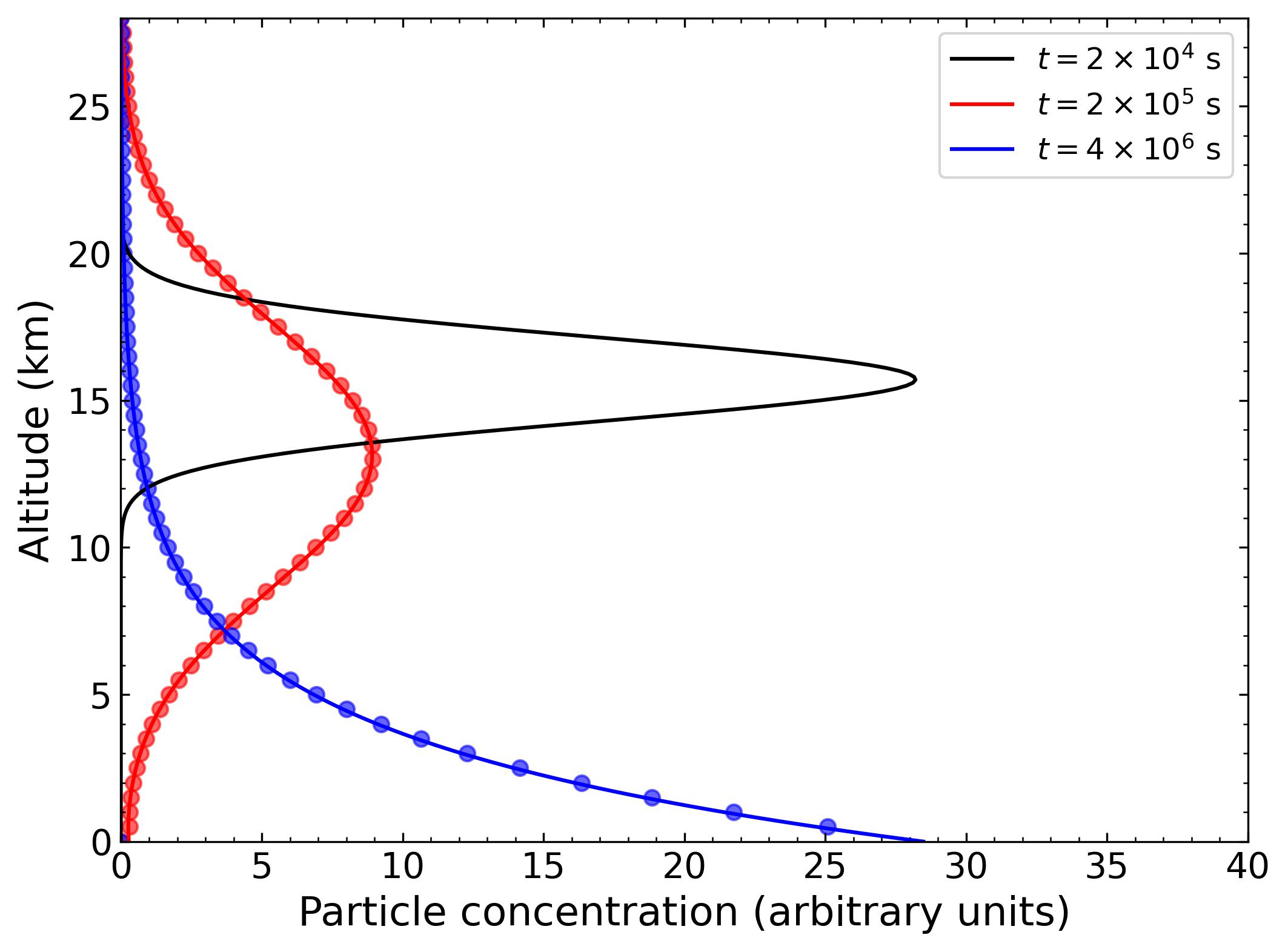} 
	\caption{Pure diffusion solution in an atmosphere with $H_\rho$= 3.5 km, assuming a constant eddy diffusion coefficient of \( K_{zz} = 50~\mathrm{m}^2/\mathrm{s} \). Solid lines correspond to the analytical solution from Eq.~\ref{eq:analytical_diff}, while dots show the numerical results for different simulated times. The distribution corresponding to the earliest time represents the initial distribution in the numerical model.}
	\label{fig:validation_diffusion}
\end{figure}

When the atmospheric density is constant, Eq.~\ref{eq:analytical_diff} is directly applicable. However, if the density follows an exponential profile, \( \rho = \rho_0 e^{-z/H_\rho} \), with constant scale height \( H_\rho \), the solution remains valid by replacing \( W \) with \( K_{zz}/H_\rho \). The excellent agreement with our model results, as illustrated in Figure~\ref{fig:validation_diffusion}, confirms the accuracy of the diffusion treatment. Please note that our solutions also show an excellent match to those of \citet{Toon1988}.
\\

\begin{figure}[h]
	\centering
	\includegraphics[width=1.\textwidth]{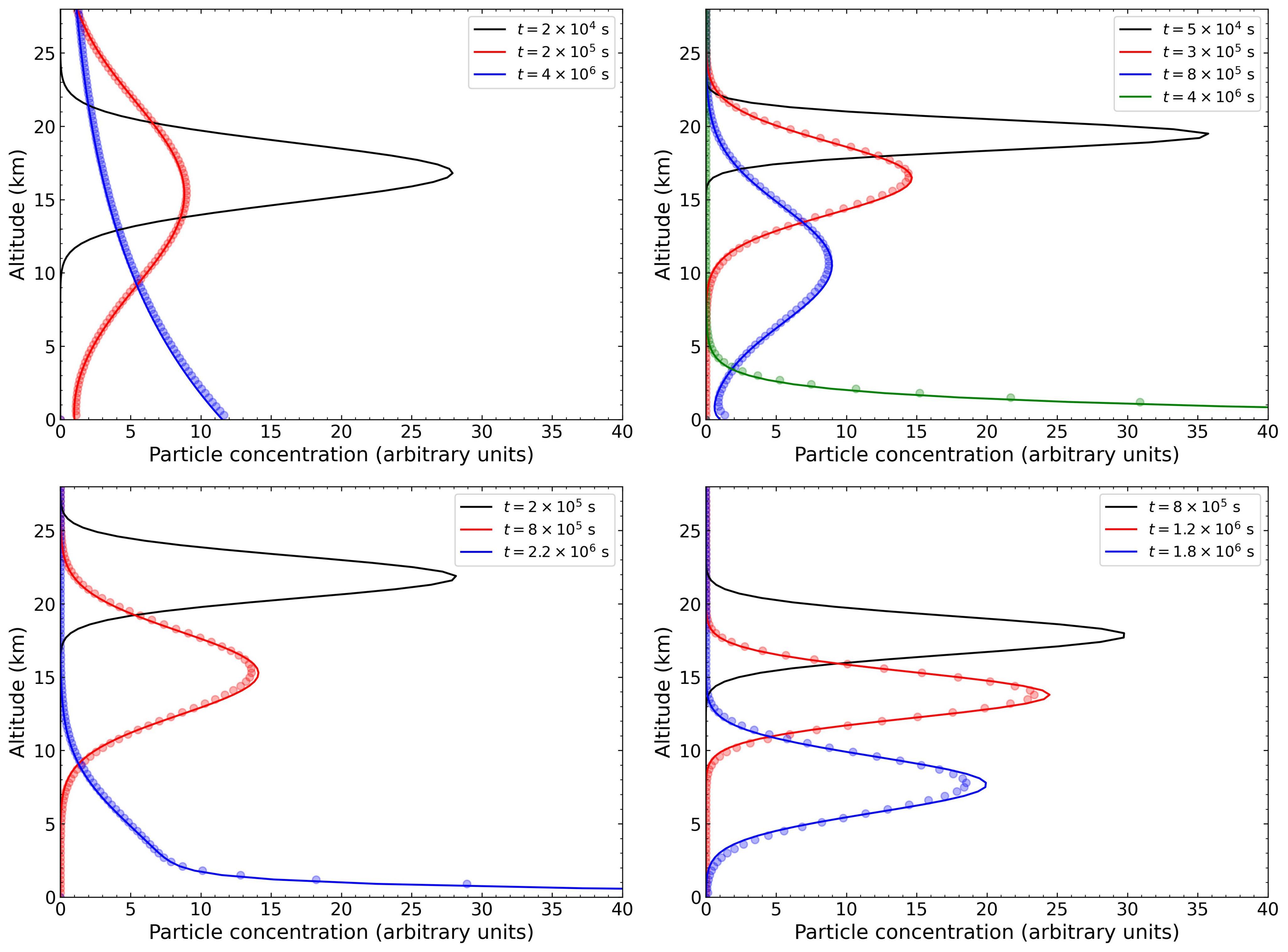} 
	\caption{Analytical solutions (solid lines) of the advection-diffusion equation in an atmosphere with $H_\rho$= 5.5 km compared with the numerical model (dots) for the following cases: a) \( W = 0.01~\mathrm{m}/\mathrm{s} \) and \( K_{zz} = 100~\mathrm{m}^2/\mathrm{s} \); b) \( W = -0.01~\mathrm{m}/\mathrm{s} \) and \( K_{zz} = 10~\mathrm{m}^2/\mathrm{s} \); c) \( W = -0.01~\mathrm{m}/\mathrm{s} \) and \( K_{zz} = 5~\mathrm{m}^2/\mathrm{s} \); d) \( W = -0.01~\mathrm{m}/\mathrm{s} \) and \( K_{zz} = 1~\mathrm{m}^2/\mathrm{s} \). The time step is \( 10^4~\mathrm{s} \). The values of \( W \) and \( K_{zz} \) were chosen to match those used in \citet{Toon1988}. In each panel, the distribution corresponding to the earliest time represents the initial distribution in the numerical model.}
	\label{fig:validation_advection-diffusion}
\end{figure}

To evaluate cases with both advection and diffusion, we follow the same procedure and use again Eq.~\ref{eq:analytical_diff}, but replacing \( W \text{ with } W + K_{zz}/H_p \). This is shown in Figure~\ref{fig:validation_advection-diffusion}, where the comparison between the analytical and numerical cases is performed for different combinations of \( W \) and \( K_{zz} \) values. The results of this comparison are again satisfactory, although a larger discrepancy is observed in the case with the lowest diffusion (\( K_{zz} = 1~\mathrm{m}^2/\mathrm{s} \)), similarly to what was reported by \citet{Toon1988}. This discrepancy arises primarily from numerical diffusion introduced by the advection scheme. While the technique proposed by \citet{Toon1988} to minimize numerical diffusion proves effective, it does not eliminate it entirely---particularly when the advection term \( W \) becomes significant relative to \( K_{zz} \). In our model, however, the lowest \( K_{zz} \) values occur near the tropopause (see Figure~\ref{fig:Kprofile}), where sedimentation velocities tend to be small for typical particle sizes (see Figure~\ref{fig:vsed_profile}), which rarely exceed radii of 3~\(\mu\)m under typical conditions, as shown in Figure \ref{fig:concentration_evolution}.  
  
\begin{figure}[h]
	\centering
	\includegraphics[width=0.75\textwidth]{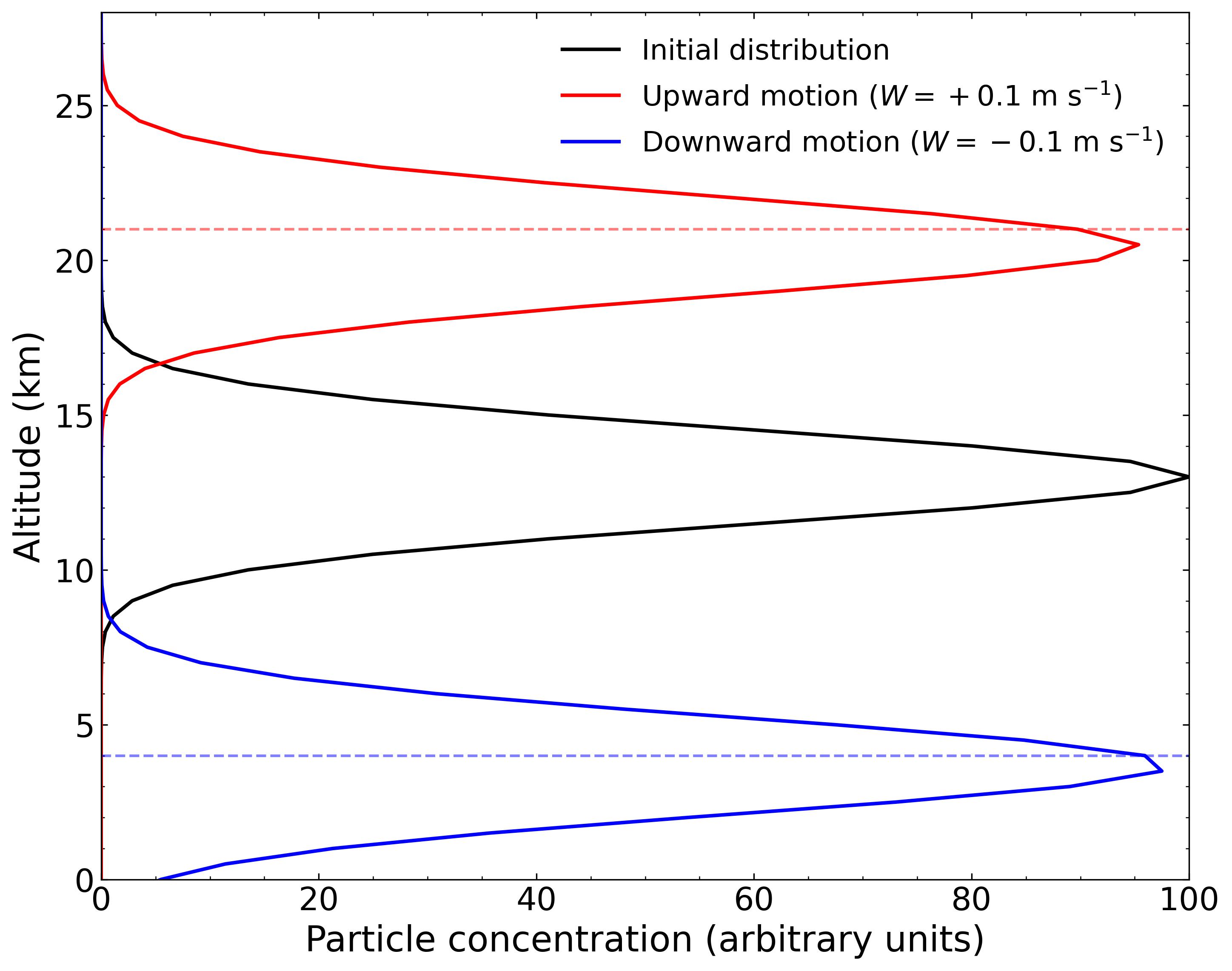} 
	\caption{ Validation of the numerical code for the case of pure advection. The red curve corresponds to upward motion and the blue one to downward motion. The dashed lines indicate where the peaks of the same-colored curves should be found. In the ideal case, the shape of the initial distribution would not change and, as shown, the peaks exhibit little distortion.}
	\label{fig:advection_validation}
\end{figure}

The validation of the pure advection case is carried out by analyzing both the displacement and the shape of the initial concentration profile, after imposing vertical velocities of equal magnitude (\( |W| = 0.1~\mathrm{m} \mathrm{s}^{-1} \)) but in opposite directions. In a purely advective scenario, where particles move vertically at a constant speed, the initial distribution should ideally shift without undergoing any distortion. Figure~\ref{fig:advection_validation} presents the outcome of this test, showing that the initial profile experiences minimal distortion. This confirms that the model accurately represents advection and that the strategy implemented to suppress numerical diffusion is effective. Minor deviations in peak position relative to the expected displacement are not considered significant, as the velocities in the  model are not known to such high precision, as pointed out by \citet{Toon1988}. Furthermore, it should be noted that pure advection does not occur in Jupiter’s atmosphere, where eddy diffusion is always present.

\begin{figure}[h]
	\centering
	\includegraphics[width=0.7\textwidth]{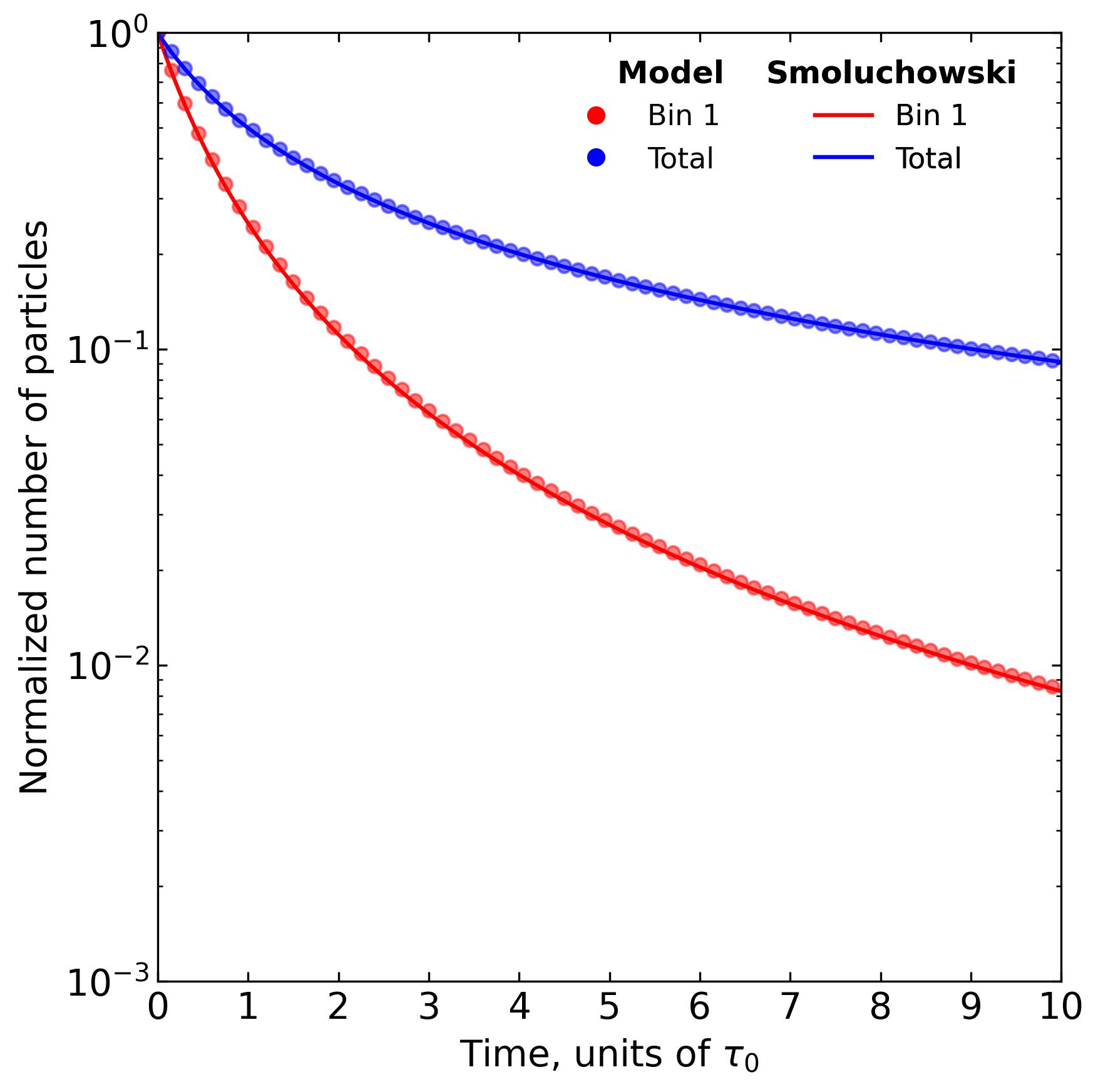} 
	\caption{Temporal evolution of the total number of particles and those remaining in the first size bin, comparing the numerical model with the analytical Smoluchowski solution for a constant coagulation kernel. Time is expressed in units of the characteristic coagulation timescale \(\tau_0 = 2/\!\left[K_{\mathrm{coag}}\,C_{\rm tot}(0)\right]\). The decrease in particle number reflects the effects of coagulation. Mass is conserved throughout the simulation. }
	\label{fig:coag_validation1}
\end{figure}

\begin{figure}[h]
	\centering
	\includegraphics[width=0.7\textwidth]{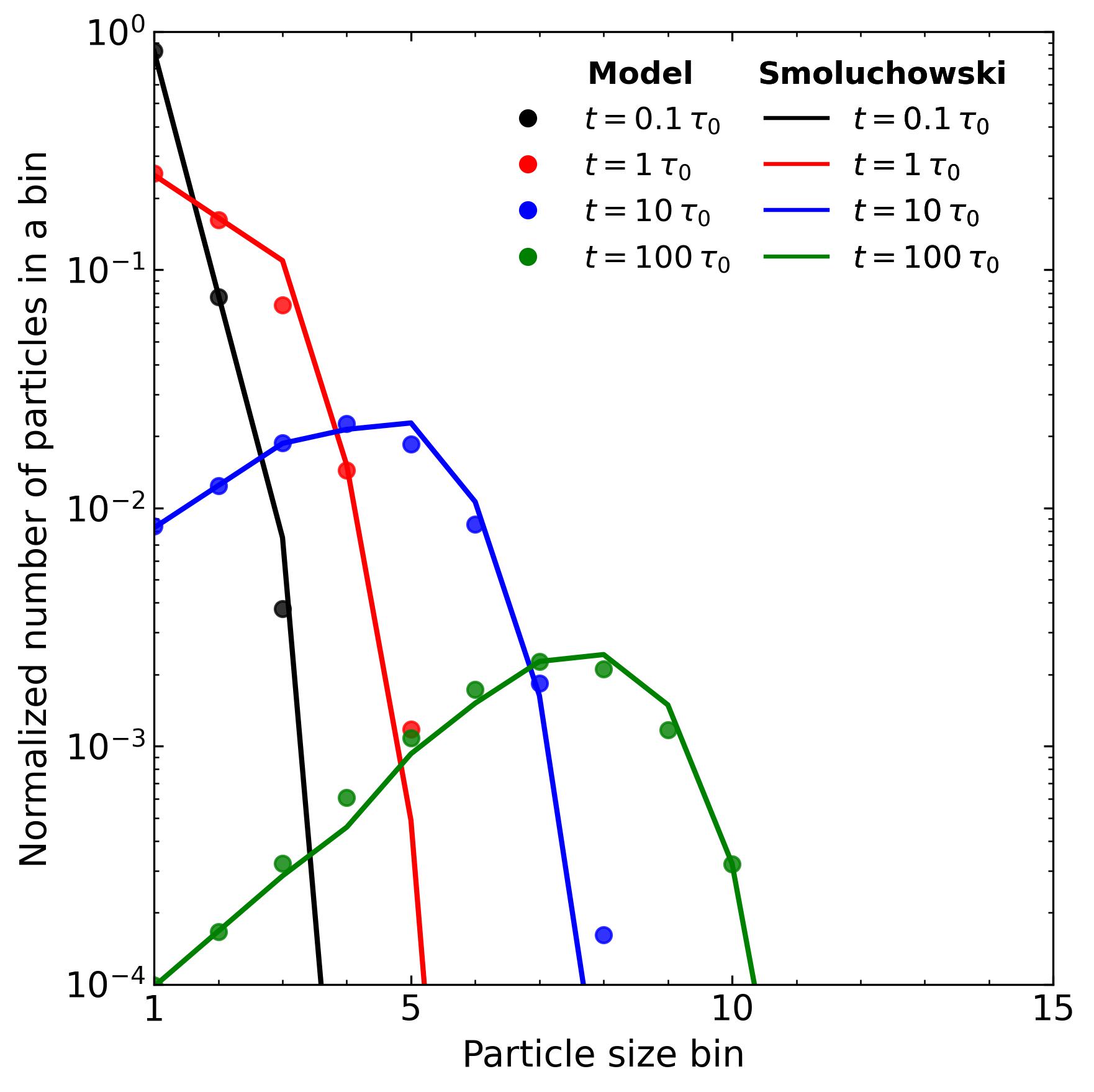} 
	\caption{ Evolution of the particle size spectrum over time, comparing the numerical model with the analytical Smoluchowski solution for a constant coagulation kernel. Times are shown in units of the characteristic coagulation timescale \(\tau_0 = 2/\!\left[K_{\mathrm{coag}}\,C_{\rm tot}(0)\right]\). Small deviations and the non-smooth Smoluchowski curves originate from translating the linearly spaced volume bins used in the analytical solution onto the model’s geometrically spaced volume bins.}
	\label{fig:coag_validation2}
\end{figure}

To evaluate the coagulation scheme, we compare the model outputs with the analytical Smoluchowski solutions for an initially monodisperse aerosol subject to a constant coagulation kernel \citep{Friedlander2000}, as introduced by \citet{Turco1979b} and later adopted by \citet{Toon1988}. Starting from a single particle in the first size bin, we track particle population evolution in terms of the characteristic coagulation time \( \tau_0 \), defined as \(\tau_0 = 2/\!\left[K_{\mathrm{coag}}\,C_{\rm tot}(0)\right]\), where \(C_{\rm tot}(0)\) is the initial total particle number concentration  \citep{Friedlander2000}. The results are shown in Figures~\ref{fig:coag_validation1} and~\ref{fig:coag_validation2}, where good agreement is observed between the model and the analytical solutions.

Figure~\ref{fig:coag_validation1} shows the temporal evolution of the total number of particles and those remaining in the first size bin. The total particle number decreases over time because multiple particles must combine to form larger aggregates. However, mass (or volume equivalently, as we assume constant particle density) is conserved throughout the simulation. Figure~\ref{fig:coag_validation2} presents the bin-population distribution at different coagulation times. Minor discrepancies between the numerical and analytical results arise from the transformation of the classical linearly spaced volume bins in the Smoluchowski solution into the geometrically increasing bins used in our model, following the procedure outlined by \citet{Turco1979b}, whose results also display similar differences. These deviations are attributable to binning artifacts rather than inaccuracies in the model, as can be inferred from the excellent agreement shown in Figure~\ref{fig:coag_validation1}, the non-smooth appearance of the analytical Smoluchowski curves in Figure~\ref{fig:coag_validation2} and by the fact that any mismatch observed at a given time does not propagate to subsequent time steps. As a final validation check, we have confirmed that the model conserves the total MCD when both vertical transport and coagulation are active. In the presence of particle injection, the MCD increases linearly in accordance with the specified injection rate.

\end{document}